\documentclass{article}

\usepackage{amsmath,amsfonts,graphicx,tikz,amsthm}
\usepackage[margin=1.5in]{geometry}
\usepackage{times}
\usepackage{bm}
\usepackage{natbib}

\newtheorem{theorem}{Theorem}

\def\T{{ \mathrm{\scriptscriptstyle T} }}

\usepackage{bbm}
\newcommand{\ind}{\mathbbm{1}}

\title{Nonparametric inference for interventional effects \\ with multiple mediators}

\author{D. BENKESER}

\begin{document}

\maketitle

\begin{abstract}
Understanding the pathways whereby an intervention has an effect on an outcome is a common scientific goal. A rich body of literature provides various decompositions of the total intervention effect into pathway specific effects. Interventional direct and indirect effects provide one such decomposition. Existing estimators of these effects are based on parametric models with confidence interval estimation facilitated via the nonparametric bootstrap. We provide theory that allows for more flexible, possibly machine learning-based, estimation techniques to be considered. In particular, we establish weak convergence results that facilitate the construction of closed-form confidence intervals and hypothesis tests. Finally, we demonstrate multiple robustness properties of the proposed estimators. Simulations show that inference based on large-sample theory has adequate small-sample performance. Our work thus provides a means of leveraging modern statistical learning techniques in estimation of interventional mediation effects.
\end{abstract}


\section{Introduction}
\label{sec1}

Recent advances in causal inference have provided rich frameworks for posing interesting scientific questions pertaining to the mediation of effects through specific biologic pathways (among others, \citet{imai2010general,valeri2013mediation,pearl2014interpretation,naimi2016mediation,zheng2017longitudinal,vanderweele2017mediation}). Foremost amongst these advances is the provision of model-free definitions of mediation parameters, which enables researchers to develop robust estimators of these quantities. The proposal of \citet{vansteelandt2017interventional} is particularly appealing. Building on the prior work of \citet{vanderweele2014effect}, the authors propose \emph{interventional mediation effects}. In contrast to other mediation effects, the proposed effects do not rely on untestable cross-world assumptions and yield a simple decomposition of the total effect into direct effects and pathway-specific effects, which holds even when the structural dependence between mediators is unknown. 

\citet{vansteelandt2017interventional} described two approaches to estimation of the effects using parametric working models for relevant nuisance parameters. In both cases, the nonparametric bootstrap was recommended for inference. A potential limitation of the proposal is that correctly specifying a parametric working model may be difficult in many settings. In these instances, we may rely on flexible estimators of nuisance parameters, for example,  based on machine learning. When such techniques are employed, the nonparametric bootstrap does not generally guarantee valid inference. This fact motivates the present work, where we develop nonparametric efficiency theory for the interventional mediation effect parameters. This theory allows us to utilize frameworks for nonparametric efficient inference to develop estimators of the quantities of interest. We propose a one-step and a targeted minimum loss-based estimator and demonstrate that under suitable regularity conditions, both estimators are nonparametric efficient amongst the class of regular asymptotically linear estimators. The estimators also enjoy a multiple robustness property, which ensures consistency of effect estimates if at least some combination of nuisance parameters are consistently estimated. Another benefit enjoyed by our estimators is the availability of closed-form confidence intervals and hypothesis tests. 

\section{Interventional Effects}
\label{sec2}

Adopting the notation of \citet{vansteelandt2017interventional}, suppose the observed data are represented as $n$ independent copies of the random variable $O = (C, A, M_1, M_2, Y) \sim P$, where $C \in \mathcal{C}$ is a vector of confounders, $A \in \{a, a^\star\}$ is a binary intervention, $M_1 \in \mathcal{M}_1$ and $M_2 \in \mathcal{M}_2$ are mediators, and $Y \in \mathcal{Y}$ is a relevant outcome. Without loss of generality, we assume $\mathcal{Y} = (0,1)$. Certain \emph{positivity assumptions} on $P$ are required for our developments. The first is that $\mbox{pr}_P\{0 < \mbox{pr}_P(A = a \mid C) < 1\} = 1$; that is, any subgroup defined by covariates $C$ that is observed with positive probability should have some chance of receiving both interventions. Secondly, we assume a positivity assumption on the distribution of the mediators. For $a_0 = a, a^\star$, we denote by $q_{a_0, M_1, M_2}(m_1, m_2 \mid c)$ the conditional density of $(M_1,M_2)$ given $A = a_0, C = c$ and assume that for $a_0 = a, a^\star$, $\mbox{pr}_P\{\mbox{inf}_{m_1, m_2} \ q_{a_0, M_1, M_2}(m_1, m_2 \mid C)\} > 0\} = 1$, where the infimum is taken over $\mathcal{M}_1 \times \mathcal{M}_2$. Our model $\mathcal{P}$ makes no assumptions about $P$ beyond these positivity conditions. However, the efficiency theory that we develop still holds under a model that makes assumptions about $\mbox{pr}_P(A \mid C)$, including the possibility that this quantity is known exactly, as in a stratified randomized trial.

To define interventional mediation effects, notation for counterfactual random variables is required. For $a_0 \in \{a, a^\star\}$, and $j = 1,2$, let $M_j(a_0)$ denote the counterfactual value for the $j$-th mediator when $A$ is set to $a_0$. Similarly, let $Y(a_0, m_1, m_2)$ denote the counterfactual outcome under an intervention that sets $A = a_0, M_1 = m_1,$ and $M_2 = m_2$. As a point of notation, when introducing quantities whose definition depends on particular components of the random variable $O$, we will use lower case letters to denote the particular value and assume that the definition at hand applies for all values in the support of that random variable. 

The total effect of intervening to set $A = a$ versus $A = a^{\star}$ is $\psi = \mathbb{E}\{Y(a, M_1(a), M_2(a))\} - \mathbb{E}\{Y(a^{\star}, M_1(a^{\star}), M_2(a^{\star}))\}$, where we use $\mathbb{E}$ to emphasize that we are taking an expectation with respect to a distribution of a counterfactual random variable. The total effect describes the difference in counterfactual outcome considering an intervention where we set $A = a$ and allow the mediators to naturally assume the value that they would under intervention $A = a$ versus an intervention where we set $A = a^{\star}$ and allow the mediators to vary accordingly. To contrast with forthcoming effects, it is useful to write the total effect in integral form. Specifically, we use $\bar{\mathbb{Q}}_{a_0}(m_1,m_2,c)$ to denote the covariate-conditional mean of the counterfactual outcome $Y(a_0, m_1, m_2)$, $\mathbb{Q}_{M_1(a_0), M_2(a_0)}(\cdot, \cdot \mid c)$ to denote the covariate-conditional bivariate cumulative distribution function of $(M_1(a_0), M_2(a_0))$, and $Q_C$ to denote the marginal distribution of $C$. The total effect can be written as \begin{align*}
	\psi &= \int\limits_\mathcal{C} \bigg\{ \ \int\limits_{\mathcal{M}_1 \times \mathcal{M}_2} \bar{\mathbb{Q}}_{a}(m_1,m_2,c) \ d\mathbb{Q}_{M_1(a), M_2(a)}(m_1, m_2 \mid c) \\
	&\hspace{0.8in} - \int\limits_{\mathcal{M}_1 \times \mathcal{M}_2} \bar{\mathbb{Q}}_{a^\star}(m_1,m_2,c) \ d\mathbb{Q}_{M_1(a^{\star}), M_2(a^{\star})}(m_1, m_2 \mid c) \bigg\} dQ_C(c) \ . 
	&\hspace{0.2in}
\end{align*}

The total effect can be decomposed into interventional direct and indirect effects. The interventional direct effect is the difference in average counterfactual outcome under two population-level interventions. The first intervention sets $A = a$, and subsequently for individuals with $C = c$ draws mediators from $\mathbb{Q}_{M_1(a^{\star}), M_2(a^{\star})}(\cdot \mid c)$. Thus, on a population level the covariate conditional distribution of mediators in this counterfactual world is the same as it would be in a population where everyone received intervention $A = a^{\star}$. This is an example of a stochastic intervention \citep{munoz2012population}. The second intervention sets $A = a^{\star}$, and subsequently allows the mediators to naturally assume the value that they would under intervention $A = a^{\star}$, so that the population level mediator distribution is again $\mathbb{Q}_{M_1(a^{\star}), M_2(a^{\star})}(\cdot \mid c)$. The interventional direct effect compares the average outcome under these two interventions, \begin{align*}
	\psi_A &= \int\limits_\mathcal{C} \int\limits_{\mathcal{M}_1 \times \mathcal{M}_2} \{\bar{\mathbb{Q}}_{a}(m_1,m_2,c) - \bar{\mathbb{Q}}_{a^\star}(m_1,m_2,c)\}d\mathbb{Q}_{M_1(a^{\star}), M_2(a^{\star})}(m_1, m_2 \mid c) dQ_C(c) \ . 
\end{align*}

For interventional indirect effects, we require definitions for the covariate-conditional distribution of each mediator, which we denote for $j = 1,2$ by $\mathbb{Q}_{M_j(a_0)}(\cdot \mid c)$. The interventional indirect effect through $M_1$ is \begin{align*}
\psi_{M_1} = \int\limits_\mathcal{C} \bigg[ \int\limits_{\mathcal{M}_2} \int\limits_{\mathcal{M}_1} \bar{\mathbb{Q}}_{a}(m_1,m_2,c) \{d\mathbb{Q}_{M_1(a)}(m_1 \mid c) - d\mathbb{Q}_{M_1(a^{\star})}(m_1 \mid c)\} d\mathbb{Q}_{M_2(a^{\star})}(m_2 \mid c)\bigg] \\ \times dQ_C(c) \ . 
\end{align*}
As with the direct effect, this effect considers two interventions. Both interventions set $A = a$. The first intervention draws mediator values independently from the marginal mediator distributions $\mathbb{Q}_{M_1(a)}(\cdot \mid c)$ and $\mathbb{Q}_{M_2(a^{\star})}(\cdot \mid c)$, while the second intervention draws mediator values independently from the marginal mediator distributions $\mathbb{Q}_{M_1(a^{\star})}(\cdot \mid c)$ and $\mathbb{Q}_{M_2(a^{\star})}(\cdot \mid c)$. The effect thus describes the average impact of shifting the population level distribution of $M_1$, while holding the population level distribution of $M_2$ fixed. The interventional indirect effect on the outcome through $M_2$ is similarly defined as \begin{align*}
\psi_{M_2} = \int\limits_\mathcal{C} \bigg[ \int\limits_{\mathcal{M}_1} \int\limits_{\mathcal{M}_2} \bar{\mathbb{Q}}_{a}(m_1,m_2,c) d\mathbb{Q}_{M_1(a^{\star})}(m_1 \mid c) \{ d\mathbb{Q}_{M_2(a)}(m_2 \mid c) - d\mathbb{Q}_{M_2(a^{\star})}(m_2 \mid c)\} \bigg] \\ \times dQ_C(c) \ . 
\end{align*}

Note that when defining interventional indirect effects, mediators are drawn \emph{independently} from marginal mediator distributions. The final effect in the decomposition essentially describes the impact of drawing the mediators from marginal rather than joint distributions. Thus, we term this effect the \emph{covariant mediator effect}, defined as \begin{align*}
\psi_{M_1, M_2} &= \int\limits_\mathcal{C} \int\limits_{\mathcal{M}_1 \times \mathcal{M}_2} \bar{\mathbb{Q}}_{a}(m_1,m_2,c) \bigg[ d\mathbb{Q}_{M_1(a), M_2(a)}(m_1, m_2 \mid c) - d\mathbb{Q}_{M_1(a) \times M_2(a)}(m_1, m_2 \mid c) \\
&\hspace{0.1in} - \{d\mathbb{Q}_{M_1(a^{\star}), M_2(a^{\star})}(m_1, m_2 \mid c) - d\mathbb{Q}_{M_1(a^{\star}) \times M_2(a^\star)}(m_1, m_2 \mid c)\} \bigg] dQ_C(c) \ ,
\end{align*}
where $d\mathbb{Q}_{M_1(a_{0}) \times M_2(a_{0})}(m_1, m_2 \mid c) = d\mathbb{Q}_{M_1(a_{0})}(m_1 \mid c) d\mathbb{Q}_{M_2(a_{0})}(m_2 \mid c)$. \citet{vansteelandt2017interventional} discuss situations where these effects are of primary interest. 

From the above definitions, we have the following effect decomposition $\psi = \psi_A + \psi_{M_1} + \psi_{M_2} + \psi_{M_1, M_2}$. We turn now to identification and efficient estimation of each of these effects.

\section{Methods}
\label{sec3}
\subsection{Identification and statistical estimation problem} \label{sec:id_and_est}

\citet{vansteelandt2017interventional} provide assumptions under which the counterfactual mean $\bar{\mathbb{Q}}_{a_0}(m_1, m_2,c)$ is identified by $\bar{Q}_{a_0}(m_1, m_2, c) = E_P(Y \mid A = a_0, M_1 = m_1, M_2 = m_2, C = c)$. The object $\bar{Q}$ is commonly referred to as the \emph{outcome regression}, since it may generally be estimated using mean regression of the outcome $Y$ onto treatment $A$, mediators $M_1$ and $M_2$, and confounders $C$. The cumulative distribution of $(M_1(a_0),M_2(a_0))$ given $C = c$ is identified by $Q_{a_0, M_1, M_2}(m_1, m_2 \mid c) = \mbox{pr}_P(M_1 \le m_1, M_2 \le m_2 \mid A = a_0, C = c)$. We assume the existence of a density $q_{a_0, M_1, M_2}$ for the mediators with respect to a dominating measure and define marginal mediator densities $q_{a_0, M_i}(m_i \mid c) = \int_{m_j \in \mathcal{M}_j} dQ_{a_0, M_1, M_2}(m_1, m_2 \mid c)$ for $i,j = 1, 2$ and $i \ne j$. We subsequently refer to these objects as \emph{marginal} mediator distributions, though they are in fact conditional on $A = a_0$ and $C$. 

We now write the identifying formula for each effect as a statistical functional of the observed data distribution. These formulas will be useful later, when we develop plug-in estimators. We begin with the total effect, which can be expressed as $P' \to \Psi(P')$, defined for each $P'$ as \begin{align*}
\Psi(P') &= \int\limits_{\mathcal{C}}\int\limits_{\mathcal{M}_1 \times \mathcal{M}_2} \bigg\{\bar{Q}_a'(m_1, m_2, c) dQ'_{a, M_1, M_2}(m_1, m_2 \mid c)  \\
&\hspace{1.2in} - \bar{Q}'_{a^{\star}}(m_1, m_2, c) dQ'_{a^{\star}, M_1, M_2}(m_1, m_2 \mid c) \bigg\} dQ'_C(c) \ ,
\end{align*}
where we use $\bar{Q}'_{a_0}$ to denote the conditional mean of $Y$ given $A = a_0, M_1, M_2, C$ that is implied by $P'$. It is convenient to introduce shorthand to define the following shorthand to the two terms comprising the integrand, \begin{align*}
\tilde{Q}_{a, M_1, M_2}(c) &= \int\limits_{\mathcal{M}_1 \times \mathcal{M}_2} \bar{Q}_a(m_1, m_2, c) dQ_{a, M_1, M_2}(m_1, m_2 \mid c) \ , \ \mbox{and} \\  
\tilde{Q}_{a^{\star}, M_1^{\star}, M_2^{\star}}(c) &= \int\limits_{\mathcal{M}_1 \times \mathcal{M}_2} \bar{Q}_{a^{\star}}(m_1, m_2, c) dQ_{a^{\star}, M_1, M_2}(m_1, m_2 \mid c) \ ,
\end{align*}
so that we can equivalently write $\Psi(P') = \int \{\tilde{Q}_{a, M_1, M_2}(c) - \tilde{Q}_{a^{\star}, M_1^{\star}, M_2^{\star}}(c)\} dQ_C'(c)$. 
The first subscript on $\tilde{Q}$ denotes the intervention, $A = a$ or $A = a^{\star}$, under which the outcome regression is evaluated. The subscript $M_1, M_2$ denotes that the outcome regression is then standardized with respect to the joint conditional distribution of $M_1, M_2$ given $A = a, C$. If instead, we standardize with respect to the joint conditional distribution given $A = a^{\star}, C$, we use the subscript $M_1^{\star}, M_2^{\star}$. 

The interventional direct effect is identified by $P' \to \Psi_A(P')$, defined for each $P'$  as \begin{equation*}
\Psi_A(P') = \int\limits_{\mathcal{C}}\int\limits_{\mathcal{M}_1 \times \mathcal{M}_2} \{\bar{Q}'_a(m_1, m_2, c) - \bar{Q}'_{a^\star}(m_1, m_2, c)\} dQ'_{a^\star, M_1, M_2}(m_1, m_2 \mid c) dQ'_{C}(c) \ .
\end{equation*}
Similarly as above, it is convenient to define \begin{align}
\tilde{Q}_{a_0, M_1^{\star}, M_2^{\star}}(c) = \int\limits_{\mathcal{M}_1 \times \mathcal{M}_2} \bar{Q}_{a_0}(m_1, m_2, c) dQ_{a^\star, M_1, M_2}(m_1, m_2 \mid c) \ ,  
\end{align}
and $\Psi_A(P')$ can also be written $\int \{\tilde{Q}'_{a, M_1^{\star}, M_2^{\star}}(c) - \tilde{Q}'_{a^{\star}, M_1^{\star}, M_2^{\star}}(c)\} dQ'_C(c).$

The interventional indirect effect through $M_1$ is identified as $P' \to \Psi_{M_1}(P')$, where \begin{align*} \label{id_indirect}
\Psi_{M_1}(P') &= \int\limits_{\mathcal{C}} \int\limits_{\mathcal{M}_1} \int\limits_{\mathcal{M}_2} \bar{Q}'_a(m_1, m_2, c) \{dQ'_{a, M_1}(m_1 \mid c) - dQ'_{a^\star, M_1}(m_1 \mid c)\} \\ &\hspace{2.5in} \times dQ'_{a^\star, M_2}(m_j \mid c) dQ'_C(c)  \ .
\end{align*}
As above, we introduce a shorthand for the inner integrals, \begin{align*}
\tilde{Q}_{a, M_1 \times M_2^{\star}}(c) &= \int\limits_{\mathcal{M}_1} \int\limits_{\mathcal{M}_2} \bar{Q}_{a}(m_1, m_2, c) dQ_{a, M_1}(m_1 \mid c) dQ_{a^{\star}, M_2}(m_2 \mid c) \ , \\
\tilde{Q}_{a, M_1^{\star} \times M_2^{\star}}(c) &= \int\limits_{\mathcal{M}_1} \int\limits_{\mathcal{M}_2} \bar{Q}_{a}(m_1, m_2, c) dQ_{a^{\star}, M_1}(m_1 \mid c) dQ_{a^{\star}, M_2}(m_2 \mid c) \ ,
\end{align*}
so we can write $\Psi_{M_1}(P') = \int_{\mathcal{C}} \{\tilde{Q}'_{a, M_1 \times M_2^{\star}}(c) - \tilde{Q}'_{a, M_1^{\star} \times M_2^{\star}}(c)\} dQ'_C(c)$. The subscript $M_1 \times M_2$ denotes that the double integral is taken with respect to the product of the marginal mediator distributions, rather than the joint mediator distribution above, while the star superscript on $M_1$ and $M_2$ still denotes whether the marginal mediator distribution is conditional on $A = a^{\star}$, as opposed to $A = a$. 

Similarly, the interventional indirect effect through $M_2$ is identified by \begin{align*} 
\Psi_{M_2}(P') &= \int\limits_{\mathcal{C}} \int\limits_{\mathcal{M}_1} \bigg[ \int\limits_{\mathcal{M}_2} \bar{Q}'_a(m_1, m_2, c) \{dQ'_{a, M_2}(m_2 \mid c) - dQ'_{a^\star, M_2}(m_2 \mid c)\} \bigg] \\ &\hspace{3.2in} \times dQ'_{a, M_1}(m_1 \mid c) dQ'_C(c)  \ ,
\end{align*}
or equivalently $\Psi_{M_2}(P') = \int_{\mathcal{C}} \{\tilde{Q}'_{a, M_1 \times M_2}(c) - \tilde{Q}'_{a, M_1 \times M_2^{\star}}(c)\}dQ'_C(c)$, where \begin{align*}
\tilde{Q}_{a, M_1 \times M_2}(c) &= \int\limits_{\mathcal{M}_1} \int\limits_{\mathcal{M}_2} \bar{Q}_{a}(m_1, m_2, c) dQ_{a, M_1}(m_1 \mid c) dQ_{a, M_2}(m_2 \mid c) \ .
\end{align*}

The interventional covariant effect is defined as the difference between the total effect and interventional direct and indirect effects, $\Psi_{M_1, M_2} = \Psi - \Psi_A - \Psi_{M_1} - \Psi_{M_2}$. 


\subsection{Efficiency theory}

In this section, we develop efficiency theory for nonparametric estimation of interventional effects. This theory centers around the efficient influence function of each parameter. The efficient influence function is important for several reasons. First, it allows us to utilize of two existing estimation frameworks, one-step estimation \citep{ibragimov1981,Bickeletal97} and targeted minimum loss-based estimation \citep{vanderLaan:Rubin06,vanderLaan:Rose11}, to generate estimators that are nonparametric efficient. That is, under suitable regularity conditions, they achieve the smallest asymptotic variance amongst all regular estimators that, when scaled by $n^{1/2}$, have an asymptotic Normal distribution. We discuss how these estimators can be implemented in section \ref{est_section}. The second important feature of the efficient influence function is that its variance equals the variance of the limit distribution of the scaled estimators. Thus, an estimate of the variance of the efficient influence function is a natural standard error estimate, which affords closed-form Wald-style confidence intervals and hypothesis tests (Section \ref{inf_section}). Finally, the efficient influence function also characterizes robustness properties of our proposed estimators (Section \ref{robust_section}). 

To introduce the efficient influence function, several additional definitions are required. For a given distribution $P' \in \mathcal{P}$, we define $g_{a_0}'(c) = \mbox{pr}_{P'}(A = a_0 \mid C = c)$, commonly referred to as a \emph{propensity score}. For $i,j=1,2$ and $i \ne j$, we introduce the following partially marginalized outcome regressions, $\tilde{Q}_{a, M_i^\star}(m_j, c) = \int \bar{Q}_a(m_1, m_2, c) dQ_{a, M_i}(m_i \mid c).$ We also introduce notation for the indicator function $\ind_a: \{a, a^{\star}\} \rightarrow \{0,1\}$ defined by $\ind_a(\tilde{a}) = 1$ if $\tilde{a} = a$ and zero otherwise. $\ind_{a^{\star}}$ is similarly defined. 

\begin{theorem} \label{thm:eif}
Under sampling from $P' \in \mathcal{P}$, the efficient influence function evaluated on a given observation $\tilde{o}$ for the total effect is \begingroup \allowdisplaybreaks \begin{align*}
D^*(P')(\tilde{o}) &= \frac{\ind_a(\tilde{a})}{g_a'(\tilde{c})} \{\tilde{y} - \tilde{Q}_{a, M_1, M_2}'(\tilde{c})\} - \frac{\ind_{a^{\star}}(\tilde{a})}{g'_{a^{\star}}(\tilde{c})} \{\tilde{y} - \tilde{Q}'_{a^\star, M_1^{\star}, M_2^{\star}}(\tilde{c})\} \\
&\hspace{0.4in} + \tilde{Q}_{a, M_1, M_2}'(\tilde{c}) - \tilde{Q}_{a^{\star}, M_1^{\star}, M_2^{\star}}'(\tilde{c}) - \Psi(P') \ . 
\end{align*}
The efficient influence function for the interventional direct effect is \begin{align*}
D_A^*(P')(\tilde{o}) &= \frac{\ind_a(\tilde{a})}{g'_a(\tilde{c})} \frac{q_{a^\star, M_1,M_2}'(\tilde{m}_1, \tilde{m}_2 \mid \tilde{c})}{q_{a, M_1,M_2}'(\tilde{m}_1, \tilde{m}_2 \mid \tilde{c})} \{\tilde{y} - \bar{Q}_a'(\tilde{m}_1, \tilde{m}_2, \tilde{c})\} \\
&\hspace{0.2in} - \frac{\ind_{a^{\star}}(\tilde{a})}{g'_{a^\star}(\tilde{c})} \{\tilde{y} - \bar{Q}_{a^\star}'(\tilde{m}_1, \tilde{m}_2, \tilde{c})\} \\
&\hspace{0.2in} + \frac{\ind_{a^{\star}}(\tilde{a})}{g'_{a^\star}(\tilde{c})} \left[\bar{Q}_a'(\tilde{m}_1, \tilde{m}_2, \tilde{c}) - \bar{Q}_{a^\star}'(\tilde{m}_1, \tilde{m}_2, \tilde{c}) - \{\tilde{Q}'_{a, M_1^\star, M_2^\star}(\tilde{c}) - \tilde{Q}_{a^{\star}, M_1^\star,M_2^\star}'(\tilde{c})\}\right] \\ 
&\hspace{0.2in} + \tilde{Q}_{a, M_1^\star,M_2^\star}'(\tilde{c}) - \tilde{Q}_{a^{\star},M_1^\star,M_2^\star}'(\tilde{c}) - \Psi_A(P') \ . 
\end{align*}
The efficient influence function for the interventional indirect effect through $M_1$ is \begin{align*}
D_{M_1}^*(P')(\tilde{o}) &= \frac{\ind_a(\tilde{a})}{g'_a(\tilde{c})} \frac{\{q_{a, M_1}'(\tilde{m}_1 \mid \tilde{c}) - q_{a^{\star}, M_1}'(\tilde{m}_1 \mid \tilde{c})\} q_{a^\star, M_2}'(\tilde{m}_2 \mid \tilde{c})}{q_{a, M_1,M_2}'(\tilde{m}_1, \tilde{m}_2 \mid \tilde{c})} \{\tilde{y} - \bar{Q}_a'(\tilde{m}_1, \tilde{m}_2, \tilde{c})\} \\
&\hspace{0.2in} + \frac{\ind_a(\tilde{a})}{g'_a(\tilde{c})} \{\tilde{Q}_{a, M_2^{\star}}'(\tilde{m}_1, \tilde{c}) - \tilde{Q}_{a, M_1 \times M_2^\star}'(\tilde{c})\} \\
&\hspace{0.2in} - \frac{\ind_{a^{\star}}(\tilde{a})}{g'_{a^{\star}}(\tilde{c})} \{\tilde{Q}_{a, M_2^{\star}}'(\tilde{m}_1, \tilde{c}) - \tilde{Q}_{a, M_1^\star \times M_2^\star}'(\tilde{c})\} \\
&\hspace{0.2in} + \frac{\ind_{a^{\star}}(\tilde{a})}{g'_{a^\star}(\tilde{c})} \left[\tilde{Q}_{a, M_1}'(\tilde{m}_2, \tilde{c}) - \tilde{Q}_{a, M_1^{\star}}'(\tilde{m}_2, \tilde{c}) - \{ \tilde{Q}_{a, M_1 \times M_2^\star}'(\tilde{c}) - \tilde{Q}_{a, M_1^{\star} \times M_2^{\star}}'(\tilde{c})\}\right] \\
&\hspace{0.2in} + \tilde{Q}_{a, M_1 \times M_2^\star}'(\tilde{c}) - \tilde{Q}_{a, M_1^\star \times M_2^\star}'(\tilde{c}) - \Psi_{M_1}(P') \ .
\end{align*}
The efficient influence function for the interventional indirect effect through $M_2$ is \begin{align*}
D_{M_2}^*(P')(\tilde{o}) &= \frac{\ind_a(\tilde{a})}{g'_a(\tilde{c})} \frac{\{q_{a, M_2}'(\tilde{m}_2 \mid \tilde{c}) - q_{a^\star, M_2}'(\tilde{m}_2 \mid \tilde{c})\}q_{a, M_1}'(\tilde{m}_1 \mid \tilde{c})}{q_{a, M_1,M_2}'(\tilde{m}_1, \tilde{m}_2 \mid \tilde{c})} \{\tilde{y} - \bar{Q}_a'(\tilde{m}_1, \tilde{m}_2, \tilde{c})\} \\
&\hspace{0.2in} + \frac{\ind_a(\tilde{a})}{g'_a(\tilde{c})} \{\tilde{Q}_{a, M_1}'(\tilde{m}_2, \tilde{c}) - \tilde{Q}_{a, M_1 \times M_2}'(\tilde{c})\} \\
&\hspace{0.2in} - \frac{\ind_{a^{\star}}(\tilde{a})}{g'_{a^{\star}}(\tilde{c})} \{\tilde{Q}_{a, M_1}'(\tilde{m}_2, \tilde{c}) - \tilde{Q}_{a, M_1 \times M_2^\star}'(\tilde{c})\} \\
&\hspace{0.2in} + \frac{\ind_{a}(\tilde{a})}{g'_{a}(\tilde{c})} \left[\tilde{Q}_{a, M_2}'(\tilde{m}_1, \tilde{c}) - \tilde{Q}_{a, M_2^{\star}}'(\tilde{m}_1, \tilde{c}) - \{ \tilde{Q}_{a, M_1 \times M_2}'(\tilde{c}) - \tilde{Q}_{a, M_1 \times M_2^{\star}}'(\tilde{c})\}\right] \\
&\hspace{0.2in} + \tilde{Q}_{a, M_1 \times M_2}'(\tilde{c}) - \tilde{Q}_{a, M_1 \times M_2^\star}'(\tilde{c}) - \Psi_{M_2}(P') \ .
\end{align*}
\endgroup
The efficient influence function for the covariant interventional effect is $D^*_{M_1, M_2} = D^* - D^*_A - D^*_{M_1} - D^*_{M_2}$.
\end{theorem}
\noindent A proof of Theorem \ref{thm:eif} is provided in the web supplement. 

\subsection{Estimators} \label{est_section}

We propose estimators of each interventional effect using one-step and targeted minimum loss-based estimation. Both techniques develop along a similar path. We first obtain estimates of the propensity score, outcome regression, and joint mediator distribution; we collectively refer to these quantities as \emph{nuisance parameters}. With estimated nuisance parameters in hand, we subsequently apply a correction based on the efficient influence function to the nuisance estimates. 

To estimate the propensity score, we can use any suitable technique for mean regression of the binary outcome $A$ onto confounders $C$. Working logistic regression models are commonly used for this purpose, though semi- and nonparametric alternatives would be more in line with our choice of model. We denote by $g_{n,a_0}(c)$ the chosen estimate of $g_{a_0}(c)$. Similarly, the outcome regression can be estimated using mean regression of the outcome $Y$ onto $A, M_1, M_2,$ and $C$. For example, if the study outcome is binary, logistic regression could again be used, though more flexible regression estimators may be preferred. As above, we denote by $\bar{Q}_{n, a_0}$ the estimated outcome regression evaluated under $A = a_0$, with $\bar{Q}_{n,a_0}(m_1, m_2, c)$ providing an estimate of $E_P(Y \mid A = a_0, M_1 = m_1, M_2 = m_2, C = c)$. To estimate the marginal cumulative distribution of $C$, we will use the empirical cumulative distribution function, which we denote by $Q_{n,C}$.

Estimation of the conditional joint distribution of the mediators is a more challenging proposition, as fewer tools are available for flexible estimation of conditional multivariate distribution functions. We describe one such option for situations where the mediators are discrete in the web supplement. This technique naturally extends to settings where mediators are continuous-valued by discretizing the support of the mediators using a fine grid. The approach entails generating an estimate $q_{n,a_0,M_1 \mid M_2}(\cdot \mid m_2, c)$ of the conditional density of $M_1$ given $A = a_0, M_2 = m_2, C = c$ and, separately, an estimate $q_{n,a_0,M_2}(\cdot \mid c)$ of the conditional density of $M_2$ given $A = a_0, C = c$. Together these estimates imply an estimate $q_{n,a_0, M_1,M_2}(m_1, m_2 \mid c)$ of the joint conditional density of $M_1$ and $M_2$ given $A$ and $C$, and an estimate $q_{n,a_0,M_1}(m_1 \mid c)$ of the marginal mediator distribution for $M_1$ given $A$ and $C$. The proposed approach allows any regression technique for a binary outcome to be used, which enables the incorporation of flexible estimation techniques, possibly based on machine learning.

Given estimates of nuisance parameters, we now illustrate one-step estimation for the interventional direct effect. One-step estimators of other effects can be generated similarly. A plug-in estimate of the conditional interventional direct effect given $C = c$ is the difference between \begin{equation} \label{cond_direct_eff_est} \begin{aligned}
	\tilde{Q}_{n, a, M_1^{\star}, M_2^{\star}}(c) &= \int\limits_{\mathcal{M}_1 \times \mathcal{M}_2} \bar{Q}_{n,a}(m_1, m_2, c) dQ_{n, a^{\star}, M_1, M_2}(m_1, m_2 \mid c) \ \mbox{and} \\
	\tilde{Q}_{n, a^{\star}, M_1^{\star}, M_2^{\star}}(c) &= \int\limits_{\mathcal{M}_1 \times \mathcal{M}_2} \bar{Q}_{n,a^{\star}}(m_1, m_2, c) dQ_{n, a^{\star}, M_1, M_2}(m_1, m_2 \mid c) \ . \\
\end{aligned}
\end{equation}
To obtain a plug-in estimate $\psi_{n,A}$ of $\psi_A$, we standardize the conditional effect estimate with respect to $Q_{n,C}$, the empirical distribution of $C$. Thus, the plug-in estimator of $\psi_A$ is $\psi_{n,A} = \int_{\mathcal{C}} \{\tilde{Q}_{n, a, M_1^{\star}, M_2^{\star}}(c) -  \tilde{Q}_{n, a^{\star}, M_1^{\star}, M_2^{\star}}(c)\} dQ_{n,C}(c)$. 

The one-step estimator is constructed by adding an efficient influence function-based correction to an initial plug-in estimate. Suppose we are given estimates of all relevant nuisance quantities and let $P_n'$ denote any probability distribution in $\mathcal{P}$ that is compatible with these estimates. The efficient influence function for $\psi_A$ under sampling from $P_n'$ is $D^*_A(P_n')$, and the one-step estimator is $\psi_{n,A,+} = \psi_{n,A} + n^{-1} \sum_{i=1}^n D^*_A(P'_n)(O_i)$. All other effect estimates are generated in this vein: estimated nuisance parameters are plugged in to the efficient influence function, the resultant function is evaluated on each observation, and the empirical average of this quantity is added to the plug-in estimator.

While one-step estimators are appealing in their simplicity, the estimators may not obey bounds on the parameter space in finite samples. For example, if the study outcome is binary, then the interventional effects each represent a difference in two probabilities and thus are bounded between -1 and 1. However, one-step estimators may fall outside of this range. This motivates estimation of these quantities using targeted minimum loss-based estimation, a framework for generating plug-in estimators. The implementation of such estimators is generally more involved than that of one-step estimators and we relegate specific details to the supplementary material.

\subsection{Large sample inference} \label{inf_section}

We now present a theorem establishing the joint weak convergence of the proposed estimators to a random variable with a multivariate normal distribution.  Because the asymptotic behavior of the one-step and targeted minimum loss estimators are equivalent, we present a single theorem. A discussion of the differences in regularity conditions required to prove the theorem for one-step versus targeted minimum loss estimation is provided in the web supplement. Let $\psi_{n,\cdot}$ denote the vector of (one-step or targeted minimum loss) estimates of $\psi_{\cdot} = (\psi_A, \psi_{M_1}, \psi_{M_2}, \psi_{M_1, M_2})^{\top}$ and let $D^*_{\cdot}(P')$ denote the vector of efficient influence functions defined by $$\tilde{o} \mapsto (D^*_A(P')(\tilde{o}), D^*_{M_1}(P')(\tilde{o}), D^*_{M_2}(P')(\tilde{o}), D^*_{M_1,M_2}(P')(\tilde{o}))^{\top} \ . $$

\begin{theorem} \label{weak_conv_thm}
Under regularity conditions explicitly stated in the web supplement, $n^{1/2}(\psi_{n,\cdot} - \psi_{\cdot}) \rightarrow_{\text{d}} \mbox{Normal}(0, \Sigma)$, where $\Sigma = \int D^*_{\cdot}(P)(o) D^*_{\cdot}(P)(o)^{\top} dP(o)$.
\end{theorem}

The regularity conditions required for Theorem \ref{weak_conv_thm} are typical of many problems in semiparametric efficiency theory and are generally satisfied by nuisance parameter estimates that (i) achieve a relatively fast rate of convergence to their true counterparts with respect to a relevant norm and (ii) satisfy relevant empirical process conditions. For details, see the web supplement. 

The covariance matrix $\Sigma$ may be estimated by the empirical covariance matrix of the vector $D^*(P_n')$ applied to the observed data, where $P_n'$ is any distribution in the model that is compatible with the estimated nuisance parameters. With the estimated covariance matrix, it is straightforward to construct Wald confidence intervals and hypothesis tests about the individual interventional effects or comparisons between them. For example, a straightforward application of the delta method would allow for a test of the null hypothesis that $\psi_{M_1} = \psi_{M_2}$. 

\subsection{Robustness properties} \label{robust_section}

As with many problems in causal inference, consistent estimation of interventional effects requires consistent estimation only of \emph{certain combinations} of nuisance parameters. To determine these combinations, we may study the stochastic properties of the efficient influence function. In particular, consider a parameter whose value under $P$ is $\tilde{\psi}$ and whose efficient influence function under sample from $P'$ can be written $\tilde{D}^*(P', \tilde{\psi}')$, where $\tilde{\psi}'$ is the value of the parameter of interest under $P'$. Then we may study the circumstances under which $\int \tilde{D}^*(P', \tilde{\psi}) dP(o) = 0$. This generally entails understanding which parameters of $P'$ must align with those parameters of $P$ to ensure that the influence function $\tilde{D}^*(P', \tilde{\psi})$ has mean zero under sampling from $P$. We present the results of this analysis in a theorem below and refer readers to the web supplement for the proof. 

\begin{theorem} \label{multiple_robustness}
Locally efficient estimators of the total effect and the intervention direct, indirect, and covariant effects are consistent for their respective target parameters if the following combinations of nuisance parameters are consistently estimated: \vspace{0.05in}

\noindent Total effect: $(\bar{Q}_{a}, \bar{Q}_{a^{\star}}, Q_{a, M_1, M_2}, Q_{a^{\star}, M_1, M_2})$ or $(g_a, g_{a^{\star}})$\vspace{0.05in}

\noindent Interventional direct effect: $(\bar{Q}_a, \bar{Q}_{a^{\star}}, g_{a^{\star}})$ or $(\bar{Q}_a,\linebreak[1] \bar{Q}_{a^{\star}},\linebreak[1] Q_{a, M_1, M_2},\linebreak[1] Q_{a^{\star}, M_1, M_2})$ or $(Q_{a, M_1, M_2},\linebreak[1] Q_{a^{\star}, M_1, M_2},\linebreak[1] g_{a^{\star}},\linebreak[1] g_{a})$; \vspace{0.05in}

\noindent Inverventional indirect effect through $M_1$: $(\bar{Q}_a, Q_{a,M_1}, Q_{a^{\star}, M_1}, Q_{a^{\star}, M_2})$ or $(g_{a},\linebreak[1] g_{a^{\star}},\linebreak[1] Q_{a,M_1, M_2},\linebreak[1] Q_{a^{\star}, M_1},\linebreak[1] Q_{a^{\star}, M_2})$ or $(\bar{Q}_a,\linebreak[1] g_a,\linebreak[1] g_{a^{\star}},\linebreak[1] Q_{a^{\star}, M_2})$ or $(\bar{Q}_a,\linebreak[1] g_a,\linebreak[1] g_{a^{\star}},\linebreak[1] Q_{a,M_1})$;\vspace{0.05in}

\noindent Inverventional indirect effect through $M_2$: $(\bar{Q}_a, Q_{a, M_2}, Q_{a^{\star}, M_2},\linebreak[1] Q_{a,M_1})$ or $(g_{a},\linebreak[1] g_{a^\star},\linebreak[1] Q_{a, M_1, M_2},\linebreak[1] Q_{a^{\star}, M_2})$ or $(\bar{Q}_{a},\linebreak[1] g_{a},\linebreak[1] g_{a^\star},\linebreak[1] Q_{a, M_1})$ or $(\bar{Q}_{a}, g_{a}, g_{a^\star}, Q_{a, M_2})$; \vspace{0.05in}

\noindent Interventional covariant effect: $(\bar{Q}_a, \bar{Q}_{a^{\star}}, Q_{a, M_1, M_2},\linebreak[1] Q_{a^{\star}, M_1, M_2})$ or $(g_a,\linebreak[1] g_a^{\star},\linebreak[1] \bar{Q}_a,\linebreak[1] \bar{Q}_{a^{\star}},\linebreak[1] Q_{a, M_1},\linebreak[1] Q_{a^{\star}, M_2})$ or $(g_a, g_{a^{\star}}, Q_{a, M_1, M_2}, Q_{a^{\star}, M_1}, Q_{a^{\star}, M_2})$.
\end{theorem}

Theorem \ref{multiple_robustness} provides sufficient, but not necessary, conditions for consistent estimation of each effect. For example, a consistent estimate of the total effect is implied by a consistent estimate of $\tilde{Q}_{a, M_1, M_2}$ and $\tilde{Q}_{a^{\star}, M_1^{\star}, M_2^{\star}}$, a condition that is generally weaker than requiring consistent estimation of the outcome regression and joint mediator distribution. Because our estimation strategy relies on estimation of the joint mediator distribution, we have described robustness properties in terms of the large sample behavior of estimators of those quantities. 

\subsection{Extensions}

Generalization to other effect scales requires only minor modifications. In particular, we can (i) determine  the portions of the efficient influence function that pertain to each component of the additive effect; (ii) develop a one-step or targeted minimum loss estimator for each component separately; (iii) use the delta method to derive the resulting influence function. In the web supplement, we illustrate an extension to a multiplicative scale.

Our results can also be extended to estimation of interventional effects for more than two mediators. As discussed in \citet{vansteelandt2017interventional}, when there are more than two mediators, say $M_1,\dots, M_t$, there are many possible path-specific effects. However, our scientific interest is usually restricted to learning effects that are mediated through each of the mediators, rather than all possible path-specific effects. Moreover, strong untestable assumptions are required to infer all path-specific effects, including assumptions about the direction of the causal effects between mediators. Therefore, it may be of greatest interest to evaluate direct and indirect effects such as \begin{align*}
	\psi_{t,A} &= \int\limits_\mathcal{C} \int\limits_{\mathcal{M}_1 \times \dots \times \mathcal{M}_t} \{\bar{\mathbb{Q}}_{a}(m_1, \dots, m_t, c) - \bar{\mathbb{Q}}_{a^\star}(m_1, \dots, m_t, c) \} \\ 
	&\hspace{1.5in} \times d\mathbb{Q}_{M_1(a^{\star}), \dots, M_t(a^{\star})}(m_1, \dots, m_t \mid c) dQ_C(c) \ ,
\end{align*}
which describes the effect of setting $A = a$ versus $A = a^\star$, while drawing all mediators from the joint conditional distribution given $A = a^{\star}, C$, and for $s = 1, \dots, t$, \begin{align*}
	\psi_{t,M_s} = \int\limits_\mathcal{C} \left[ \int\limits_{\mathcal{M}_1 \times \dots \times \mathcal{M}_t} \bar{\mathbb{Q}}_a(m_1, m_2, c) \{d\mathbb{Q}_{M_s(a)}(m_s \mid c) - d\mathbb{Q}_{M_s(a^{\star})}(m_s \mid c)\} \right. \\ \left.
	\times \prod_{u=1}^{s-1} d\mathbb{Q}_{M_u(a)}(m_u \mid c) \prod_{v=s+1}^{t}  d\mathbb{Q}_{M_v(a^{\star})}(m_v \mid c)\right]dQ_C(c) \ ,
\end{align*}
which describes the effect of setting $M_s$ to the value it would assume under $A = a$ versus $A = a^{\star}$ while drawing $M_1, \dots, M_{s-1}$ from their respective marginal distributions given $A = a, C$ and drawing $M_{s+1}, \dots, M_{t}$ from their marginal distribution given $A = a^\star, C$. We provide relevant efficiency theory for these parameters in the web supplement.

\section{Simulation}

We evaluated the small sample performance of our estimators via Monte Carlo simulation. Data were generated as follows. We simulated $C$ by drawing two random variables, $(C_1, C_2)$, independently from a Uniform(0,1) distribution. The treatment variable $A$ was, given $C = c$, was drawn from a Bernoulli distribution with $g_{a}(c) = \mbox{logit}^{-1}(-1 + c_1 + c_2)$ and $g_{a^{\star}}(c) = 1 - g_a(c)$. Here, we consider $a = 1$ and $a^{\star} = 0$. Given $C = c, A = a_0$, the first mediator $M_1$ was generated by taking draws from a geometric distribution with success probability $\mbox{logit}^{-1}(-1 + 0.25 c_1 + 0.25 a_0)$. Any draw of five or greater was set equal to five. The second mediator was generated from a similarly truncated geometric distribution with success probability $\mbox{logit}^{-1}(-1 + 0.25 c_1 + 0.35 a_0)$. Given $C = c, A = a_0, M_1 = m_1, M_2 = m_2$, the outcome $Y$ was drawn from a Bernoulli distribution with success probability $\mbox{logit}^{-1}(-1 + c_1 - c_2 + 0.5 m_1 + 0.5 m_2 + a_0)$. The true total effect is approximately 0.10, which decomposes into a direct effect of 0.15, an indirect effect through $M_1$ of -0.02, an indirect effect through $M_2$ of -0.03 and a covariant effect of 0. 

The nuisance parameters were estimated using regression stacking \citep{Wolpert92,Breiman96d}, also known as super learning \citep{vanderLaan:Polley:Hubbard07} using the \texttt{SuperLearner} package for the \texttt{R} language \citep{Polley:vanderLaan13}. We used this package to generate an ensemble of a main-terms logistic regression (as implemented in the \texttt{SL.glm} function in \texttt{SuperLearner}), polynomial multivariate adaptive regression splines (\texttt{SL.earth}), and a random forest (\texttt{SL.ranger}). The ensemble was built by selecting the convex combination of these three estimators that minimized ten-fold cross-validated deviance. 

We evaluated our proposed estimators under this data generating process at sample sizes of $250, 500, 1,000,$ and $2,000$. At each sample size, we simulated 1,000 data sets. Point estimates were compared in terms of their Monte Carlo bias, standard deviation, mean squared error. We evaluated weak convergence by visualizing the sampling distribution of the estimators after centering at the true parameter value and scaling by an oracle standard error, computed as the Monte Carlo standard deviation of the estimates, as well as scaling by an estimated standard error based on the estimated variance of the efficient influence function. Similarly, we evaluated the coverage probability of a nominal 95\% Wald-style confidence interval based on the oracle and estimated standard errors.

\begin{figure}
\centering
\includegraphics[width=0.8\textwidth]{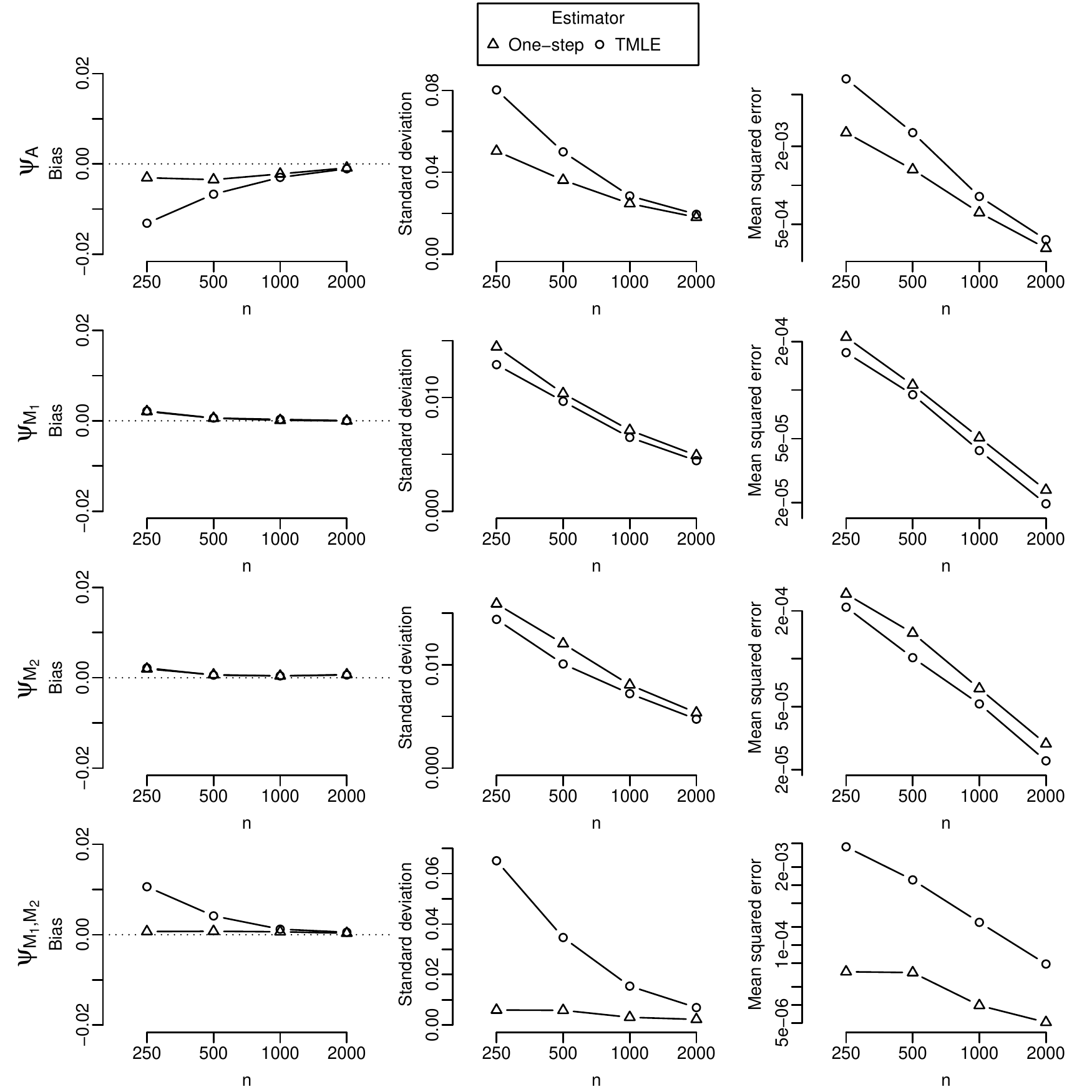}
\caption{Comparison of one-step and targeted minimum loss estimators (TMLE) in terms of their Monte Carlo-estimated bias, standard deviation, and mean squared-error for the interventional direct ($\psi_A$), indirect ($\psi_{M_1}, \psi_{M_2}$), and covariant ($\psi_{M_1,M_2}$ effects.}
\label{fig:estimation}
\end{figure}

In terms of estimation, one-step and targeted minimum loss estimators behave as expected in large samples (Figure \ref{fig:estimation}). The estimators are approximately unbiased in large samples and have mean squared error appropriately decreasing with sample size. Comparing the two estimation strategies, we see that one-step tended to perform better than targeted minimum loss estimation for the interventional direct and covariant effects. Further examination of the results revealed that the second-stage model fitting required by the targeted minimum loss approach was unstable in small samples, leading to extreme results in some data sets. 

\begin{figure}
\centering
\includegraphics[width=0.8\textwidth]{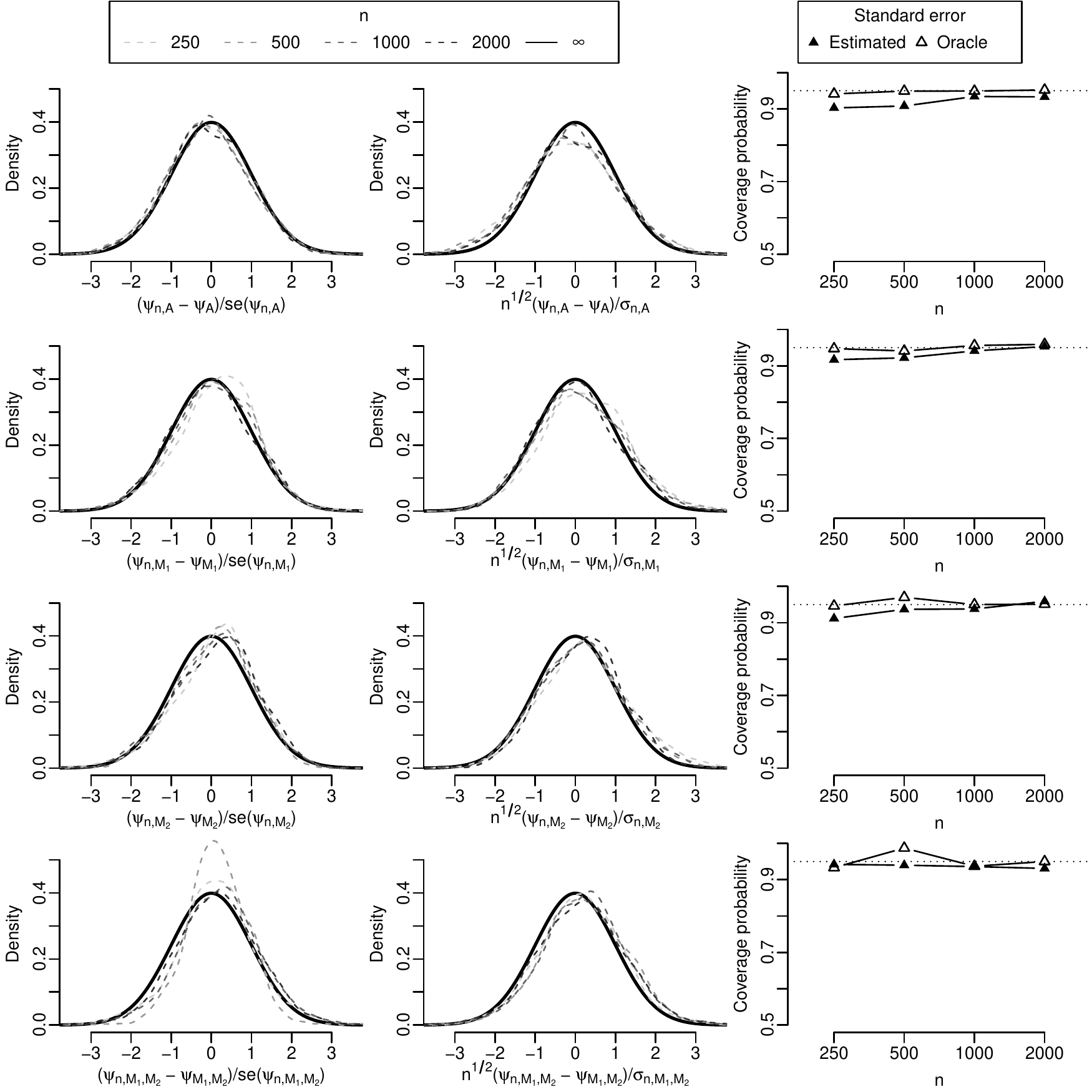}
\caption{Illustration of weak convergence and Wald-style confidence intervals based on the one-step estimator. The left two columns show the kernel density estimate of the sampling distribution of the centered estimates of interventional effects scaled by the oracle standard error (left) and by their estimated standard error (middle). In each case, the asymptotic distribution is shown in black. The right panel shows coverage probability of a nominal 95\% Wald-style confidence interval based on an oracle standard error (solid triangle) and an estimated standard error (open triangle).}
\label{fig:inf_aiptw}
\end{figure}

\begin{figure}
\centering
\includegraphics[width=0.8\textwidth]{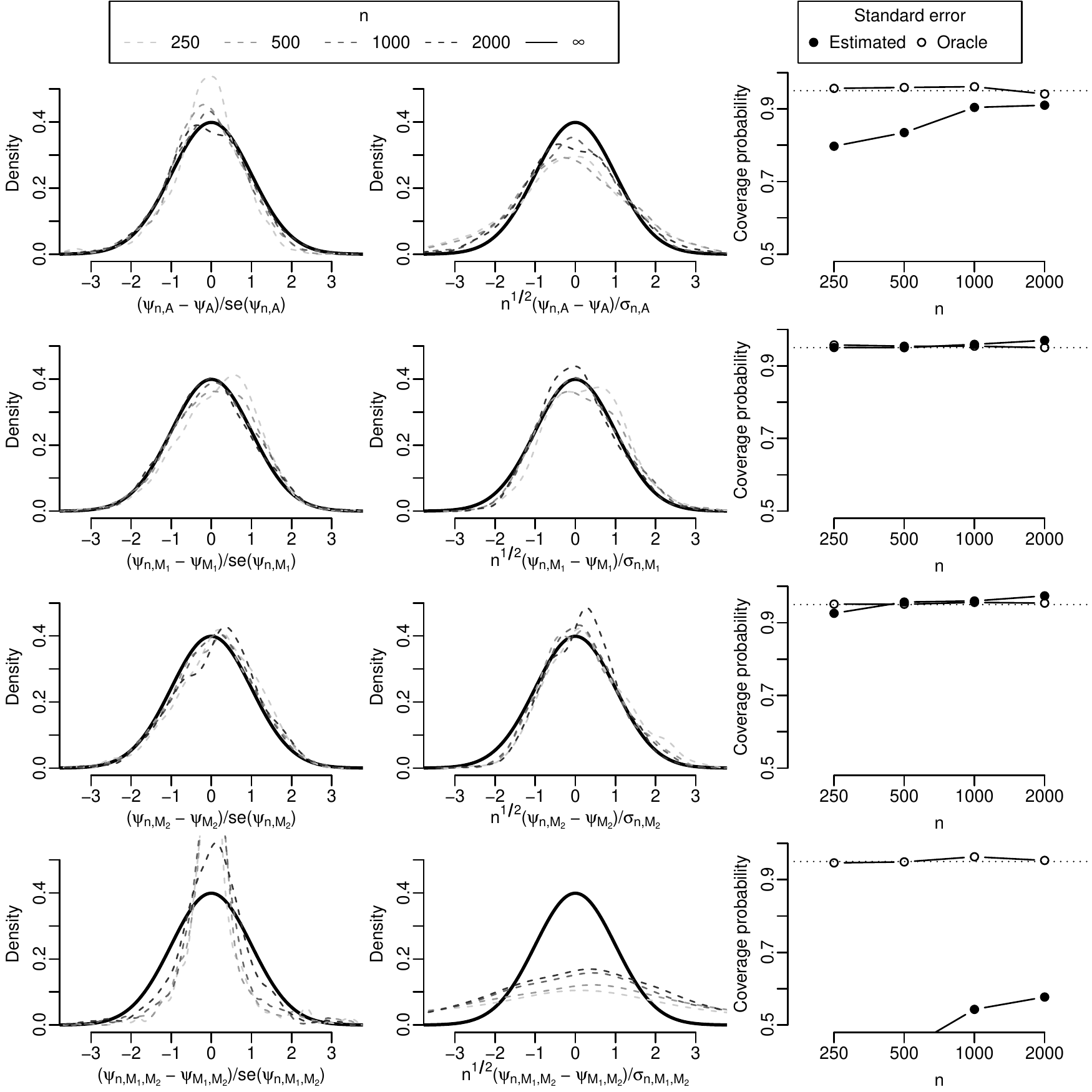}
\caption{Illustration of weak convergence and Wald-style confidence intervals based on the targeted minimum loss estimator. The left two columns show the kernel density estimate of the sampling distribution of the centered estimates of interventional effects scaled by the oracle standard error (left) and by their estimated standard error (middle). In each case, the asymptotic distribution is shown in black. The right panel shows coverage probability of a nominal 95\% Wald-style confidence interval based on an oracle standard error (solid circle) and an estimated standard error (open circle).}
\label{fig:inf_tmle}
\end{figure}

The sampling distribution of the centered and scaled one-step estimator was approximately a standard normal distribution (Figure \ref{fig:inf_aiptw}), as predicted by our theory. Confidence intervals based on an oracle standard error came close to nominal coverage in all sample sizes, while those based on an estimated standard error tended to have marginal under-coverage in small samples, but never worse than 90\%. For the targeted minimum loss estimators, we found that the indirect effect estimators behaved as expected (Figure \ref{fig:inf_tmle}). However, the instability in the second stage fitting for the direct and indirect effects led to poor results in small samples. Nevertheless, we do see evidence that in large samples these estimators begin to behave as expected. 

\section{Discussion}

The behavior of the direct effect targeted minimum loss estimator in the simulation is surprising as generally we expect comparable or better performance of such estimators relative to one-step estimators. We explored whether the poor performance was due to the targeting procedure over-fitting the conditional effect parameter in small samples by implementing a uniformly least favorable submodel \citep{ULFM16}, which should entail the minimum amount of additional model fitting required to satisfy the requisite efficient influence function estimating equation. However, the results were largely the same. It may be that the poor behavior is due to the fact that our targeted minimum loss procedure does not yield a compatible plug-in estimator of the vector $\psi_{\cdot}$, in the sense that there is likely no distribution $P_n'$ that is compatible with all of the various nuisance estimators after the second-stage model fitting. A more parsimonious approach could consider a uniformly least favorable submodel that simultaneously targets the joint mediator density and outcome regression. We leave to future work an implementation of such an estimator and hypothesize that it may improve small-sample performance.

\citet{zheng2017longitudinal} proposed estimators of longitudinal mediation effects that do not require estimates of conditional mediator distributions. These quantities can be avoided due to the fact that (i) the mediation effects of interest can be represented as functionals of iteratively-defined conditional means that can be estimated using sequential regression and (ii) the efficient influence function of the target parameters can be written in terms of nuisance parameters that do not involve mediator distributions. These techniques could be directly applied to generate estimators of interventional direct effect; however, application to indirect and covariant effects seems a more challenging proposition. The difficulty arises because interventional indirect effects require marginalizing the outcome regression relative to the product of marginal mediator distributions, while the direct effect and those effects studied in \citet{zheng2017longitudinal} require marginalization with respect to the joint distribution. In future work, we will explore whether modifications of the \citet{zheng2017longitudinal} approach can be made for interventional effects. 

An \texttt{R} package \texttt{intermed} with implementations of the proposed methods is available in the web supplementary material.

\bibliographystyle{apalike}
\bibliography{refs}

\clearpage

\section*{Supplementary material}

\section{Proof Theorem 1}

To find the efficient influence function, we assume that $O$ is discrete and find the efficient influence function of the nonparametric maximum likelihood estimate. In the end, these derivations do not rely on the discreteness of $O$, so we can conclude that the resultant influence function is the efficient influence function. The derivations unfold by writing each effect as some mapping $\hat{\Psi}: \mathcal{P} \rightarrow (-1, 1)$. Thus, the nonparametric maximum likelihood estimator can be written as $\hat{\Psi}(P_n)$, where $P_n$ is the empirical measure of $O_1, \dots, O_n$, and the true parameter can be written as $\hat{\Psi}(P)$. Moreover, we can represent the estimator $\hat{\Psi}^*(\mathcal{F}_n)$ where $\mathcal{F}_n$ is a vector of empirical means of the form $n^{-1} \sum_{i=1}^n \ind_{\cdot}(O_i)$. In the proof, we will make use of the shorthand notation $P f = \int f(o) dP(o)$, for any $P$-integrable function $f$. Similarly, we will write $P_n f = \int f(o) dP_n(o) = n^{-1}\sum_{i=1}^n f(O_i)$. 

\subsection{Total effect} The proof of the efficient influence function for the total effect has been presented many times in the literature (e.g., \citet{vanderLaan:Rose11}). 

\subsection{Interventional direct effect} The interventional direct effect writes as \begin{equation*}
\hat{\Psi}_A(P) = \sum\limits_{y,m_1,m_2,c} \left\{ y \left(\frac{P \ind_{y, a, m_1, m_2, c}}{P \ind_{a,m_1,m_2,c}} - \frac{P \ind_{y, a^{\star}, m_1, m_2, c}}{P \ind_{a^{\star},m_1,m_2,c}} \right) \frac{P \ind_{a^{\star},m_1, m_2, c}}{P \ind_{a^{\star}, c}} P \ind_c \right\} \ , 
\end{equation*}
where $\ind_{y,m_1,m_2,c}(\tilde{o}) = \ind_y(\tilde{y})\ind_{m_1}(\tilde{m}_1)\ind_{m_2}(\tilde{m}_2)\ind_{c}(\tilde{c})$ and other indicator functions are similarly defined. Note then that $P\ind_{y,m_1,m_2,c} = P(Y = y, M_1 = m_1, M_2 = m_2, C = c)$ represents a parameter of the distribution $P$. Because the data are discrete, the delta method implies that the efficient score for this parameter evaluated on an observation $\tilde{o}$ is \[
\frac{d\hat{\Psi}_A(P)}{dP\ind_{y,m_1,m_2,c}}(\tilde{o}) \{ \ind_{y,m_1,m_2,c}(\tilde{o}) - P \ind_{y,m_1,m_2,c}\} \ , 
\] and similarly for other parameters of the form $P \ind_{\cdot}$. We thus define the efficient score for each in turn and add them together to give the form of the efficient influence function. Given $\tilde{o}$, \begingroup \allowdisplaybreaks \begin{align*}
&\sum\limits_{y,m_1,m_2,c} \frac{\partial \hat{\Psi}(P)}{\partial P\ind_{y, a, m_1, m_2, c}}(\tilde{o}) \left\{\ind_{y, a, m_1, m_2, c}(\tilde{o}) - P \ind_{y, a, m_1, m_2, c} \right\} \\
&\hspace{0.5in} = \tilde{y}\left\{\frac{\ind_a(\tilde{a})}{g_a(\tilde{c})} \frac{q_{a^\star,M_1,M_2}(\tilde{m}_1, \tilde{m}_2 \mid \tilde{c})}{q_{a,M_1, M_2}(\tilde{m}_1, \tilde{m}_2 \mid \tilde{c})} \right\} - \int\limits_{\mathcal{C}} \tilde{Q}_{a, M_1^{\star}, M_2^{\star}}(c) dQ_C(c) \ , \\
&\sum\limits_{m_1,m_2,c} \frac{\partial \hat{\Psi}(P)}{\partial P\ind_{a, m_1, m_2, c}}(\tilde{o}) \left\{\ind_{a, m_1, m_2, c}(\tilde{o}) - P \ind_{a, m_1, m_2, c} \right\} \\
&\hspace{0.5in} =  -\bar{Q}_a(\tilde{m}_1, \tilde{m}_2, \tilde{c}) \left\{\frac{\ind_a(\tilde{a})}{g_a(\tilde{c})} \frac{q_{a^\star,M_1,M_2}(\tilde{m}_1, \tilde{m}_2 \mid c)}{q_{a,M_1, M_2}(\tilde{m}_1, \tilde{m}_2 \mid \tilde{c})} \right\}  + \int\limits_{\mathcal{C}} \tilde{Q}_{a,M_1^{\star},M_2^{\star}}(c) dQ_C(c) \ , \\
&\sum\limits_{y,m_1,m_2,c} \frac{\partial \hat{\Psi}(P)}{\partial P\ind_{y, a^{\star}, m_1, m_2, c}}(\tilde{o}) \left\{\ind_{y, a^{\star}, m_1, m_2, c}(\tilde{o}) - P \ind_{y, a^{\star}, m_1, m_2, c} \right\} \\
&\hspace{0.5in} = -\tilde{y} \frac{\ind_{a^{\star}}(\tilde{a})}{g_{a^{\star}}(\tilde{c})} + \int\limits_{\mathcal{C}} \tilde{Q}_{a^{\star}, M_1^{\star},M_2^{\star}}(c) dQ_C(c) \ , \\
&\sum\limits_{m_1,m_2,c} \frac{\partial \hat{\Psi}(P)}{\partial P\ind_{a^{\star}, m_1, m_2, c}}(\tilde{o}) \left\{\ind_{a^{\star}, m_1, m_2, c}(\tilde{o}) - P \ind_{a^{\star}, m_1, m_2, c} \right\} \\
&\hspace{0.5in} =  \bar{Q}_a(\tilde{m}_1, \tilde{m}_2, \tilde{c}) \frac{\ind_{a^{\star}}(\tilde{a})}{g_{a^{\star}}(\tilde{c})} - \int\limits_{\mathcal{C}} \tilde{Q}_{a, M_1^{\star}, M_2^{\star}}(c) dQ_C(c) \ , \\
&\sum\limits_{c} \frac{\partial \hat{\Psi}(P)}{\partial P\ind_{a^{\star}, c}}(\tilde{o}) \left\{\ind_{a^{\star}, c}(\tilde{o}) - P \ind_{a^{\star}, c} \right\} = - \frac{\ind_{a^{\star}}(\tilde{a})}{g_{a^{\star}}(\tilde{c})} \{\tilde{Q}_{a, M_1^{\star}, M_2^{\star}}(\tilde{c}) - \tilde{Q}_{a^{\star}, M_1^{\star}, M_2^{\star}}(\tilde{c}) \} +  \Psi_A(P), \ \mbox{and} \\
&\sum\limits_{c} \frac{\partial \hat{\Psi}(P)}{\partial P\ind_{c}}(\tilde{o}) \left\{\ind_{c}(\tilde{o}) - P \ind_{c} \right\} = \tilde{Q}_{a, M_1^{\star}, M_2^{\star}}(\tilde{c}) - \tilde{Q}_{a^{\star}, M_1^{\star}, M_2^{\star}}(\tilde{c}) - \Psi_A(P) \ . 
\end{align*} \endgroup

\noindent These terms can be combined to arrive at the final result.

\subsection{Interventional indirect effect}
Note that the interventional indirect effect through $M_1$ writes as \[
\hat{\Psi}_{M_1}(P) = \sum\limits_{y,m_1,m_2,c} \left\{ y \frac{P \ind_{y, a, m_1, m_2, c}}{P \ind_{a,m_1,m_2,c}} \left( \frac{P \ind_{a, m_1, c}}{P \ind_{a, c}} - \frac{P \ind_{a^{\star}, m_1, c}}{P \ind_{a^{\star}, c}} \right) \frac{P \ind_{a^{\star}, m_2, c}}{P \ind_{a,c}^{\star}} P \ind_c \right\} \ . 
\]
The efficient scores for these indexing parameters are \begingroup \allowdisplaybreaks \begin{align*}
&\sum\limits_{y,m_1,m_2,c} \frac{\partial \hat{\Psi}(P)}{\partial P\ind_{y, a, m_1, m_2, c}}(\tilde{o}) \left\{\ind_{y, a, m_1, m_2, c}(\tilde{o}) - P \ind_{y, a, m_1, m_2, c} \right\} \\
&\hspace{0.2in} = \frac{\ind_a(\tilde{a})}{g_a(\tilde{c})} \frac{\{q_{a, M_1}(\tilde{m}_1 \mid \tilde{c}) - dQ_{a^\star, M_1}(\tilde{m}_1 \mid \tilde{c})\}dQ_{a^\star, M_2}(\tilde{m}_2 \mid \tilde{c})}{q_{a, M_1,M_2}(\tilde{m}_1, \tilde{m}_2 \mid \tilde{c})} \tilde{y} - \Psi_{M_1}(P) \\ 
&\sum\limits_{m_1,m_2,c} \frac{\partial \hat{\Psi}(P)}{\partial P\ind_{a, m_1, m_2, c}}(\tilde{o}) \left\{\ind_{a, m_1, m_2, c}(\tilde{o}) - P \ind_{a, m_1, m_2, c} \right\} \\
&\hspace{0.2in} = -\frac{\ind_a(\tilde{a})}{g_a(\tilde{c})} \frac{\{dQ_{a, M_1}(\tilde{m}_1 \mid \tilde{c}) - dQ_{a^\star, M_1}(\tilde{m}_1 \mid \tilde{c})\}dQ_{a^\star, M_2}(\tilde{m}_2 \mid \tilde{c})}{q_{a, M_1,M_2}(\tilde{m}_1, \tilde{m}_2 \mid \tilde{c})} \bar{Q}_a(\tilde{m}_1,\tilde{m}_2,\tilde{c}) + \Psi_{M_1}(P) \\
&\sum\limits_{m_1,c} \frac{\partial \hat{\Psi}(P)}{\partial P\ind_{a, m_1, c}}(\tilde{o}) \left\{\ind_{a, m_1, c}(\tilde{o}) - P \ind_{a, m_1, c} \right\} = \frac{\ind_a(\tilde{a})}{g_a(\tilde{c})} \tilde{Q}_{a,M_2^{\star}}(\tilde{m}_1, \tilde{c}) - \int \tilde{Q}_{a, M_1 \times M_2^\star}(c) dQ_C(c) \\
&\sum\limits_{c} \frac{\partial \hat{\Psi}(P)}{\partial P\ind_{a, c}}(\tilde{o}) \left\{\ind_{a, c}(\tilde{o}) - P \ind_{a, c} \right\} = -\frac{\ind_a(\tilde{a})}{g_a(\tilde{c})} \tilde{Q}_{a, M_1 \times M_2^\star}(\tilde{c}) + \int \tilde{Q}_{a, M_1 \times M_2^\star}(c) dQ_C(c) \\
&\sum\limits_{m_1,c} \frac{\partial \hat{\Psi}(P)}{\partial P\ind_{a^\star, m_1, c}}(\tilde{o}) \left\{\ind_{a^\star, m_1, c}(\tilde{o}) - P \ind_{a^\star, m_1, c} \right\} = -\frac{\ind_{a^{\star}}(\tilde{a})}{g_{a^\star}(\tilde{c})} \tilde{Q}_{a,M_2^\star}(\tilde{m}_1, \tilde{c}) + \int \tilde{Q}_{M_1^\star \times M_2^\star}(a, c) dQ_C(c) \\
&\sum\limits_{c} \frac{\partial \hat{\Psi}(P)}{\partial P\ind_{a^\star, c}}(\tilde{o}) \left\{\ind_{a^{\star}, c}(\tilde{o}) - P \ind_{a^\star, c} \right\} = - \frac{\ind_{a^{\star}}(\tilde{a})}{g_{a^\star}(\tilde{c})} \tilde{Q}_{a, M_1 \times M_2^\star}(\tilde{c}) + \int \tilde{Q}_{a, M_1 \times M_2^\star}(c) dQ_C(c) \\
&\hspace{2.4in} + 2 \frac{\ind_{a^{\star}}(\tilde{a})}{g_{a^\star}(\tilde{c})} \tilde{Q}_{a, M_1^\star \times M_2^\star}(\tilde{c}) - 2 \int \tilde{Q}_{a,M_1^\star \times M_2^\star}(c) dQ_C(c) \\
&\sum\limits_{m_2,c} \frac{\partial \hat{\Psi}(P)}{\partial P\ind_{a^\star, m_2, c}}(\tilde{o}) \left\{\ind_{a^\star, m_2, c}(\tilde{o}) - P \ind_{a^\star, m_2, c} \right\} = \frac{\ind_{a^{\star}}(\tilde{a})}{g_{a^\star}(\tilde{c})} \{\tilde{Q}_{a,M_1}(\tilde{m}_2, \tilde{c}) - \tilde{Q}_{a, M_1^\star}(\tilde{m}_2, \tilde{c})\} - \Psi_{M_1}(P) \\
&\sum\limits_{c} \frac{\partial \hat{\Psi}(P)}{\partial P\ind_{c}}(\tilde{o}) \left\{\ind_{c}(\tilde{o}) - P \ind_{c} \right\} = \{\tilde{Q}_{a, M_1\times M_2^{\star}}(\tilde{c}) - \tilde{Q}_{a, M_1^\star \times M_2^{\star}}(\tilde{c})\} - \Psi_{M_1}(P) 
\end{align*}
\endgroup 

\noindent These terms can be combined to arrive at the final result. The proof for the efficient influence function of $\Psi_{M_2}$ is essentially the same as above. Owing to the effect decomposition, the delta method implies that the efficient influence function of $\Psi_{M_1,M_2}$ is equal to the difference between the efficient influence function for the total effect minus the sum of those for the direct and indirect effects. 

\section{Proof of multiple robustness}

Below we use the following shorthand notation: for a given (appropriately measurable) function $f$, we define $Q_{M_1} f = \int f(o) dQ_{a, M_1}(m_1 \mid c)$, $Q_{M_j^\star} f = \int f(o) dQ_{a^\star, M_j}(m_j \mid c)$ for $j = 1,2$, and $Q_{M_1,M_2} f = \int f(o) dQ_{a, M_1,M_2}(m_1, m_2 \mid c)$. We add an apostrophe to denote these same expressions but considering sampling from $P'$ rather than $P$. The equalities shown below can be arrived at via straightforward, but extensive, algebra. For brevity, we have omitted this algebra, as it is not particularly informative.

\subsection{Total effect} Robustness of the total effect has been shown in many previous studies, e.g., \citet{vanderLaan:Rose11}. 

\subsection{Interventional direct effect} We can write $D_{A}$ as an estimating function $D_A(P, \psi_A)$ and \begingroup \allowdisplaybreaks \begin{align*}
  P \{D_A(P', \psi_A)\} &= P \left[ \frac{g_a}{g_a'} \left(Q_{M_1,M_2}' - Q_{M_1, M_2}\right) \left\{\frac{q_{a^\star, M_1, M_2}'}{q_{a, M_1, M_2}q_{a,M_1,M_2}'}\left( \bar{Q}_a' - \bar{Q}_a\right) \right\} \right] \\
  &\hspace{0.1in} + P\left\{ \left( \frac{g_a' - g_a}{g_a'} \right) Q_{M_1^{\star}, M_2^{\star}}\left( \bar{Q}_a' - \bar{Q}_a \right)  \right\} \\
  &\hspace{0.15in} + P \left\{ \frac{g_a}{g'_a} \left(Q_{M_1^{\star}, M_2^{\star}}' - Q_{M_1^{\star}, M_2^{\star}} \right) \left( \bar{Q}_a' - \bar{Q}_a \right) \right\} \\
  &\hspace{0.2in} + P \left\{ \left(\frac{g_{a^{\star}}' - g_{a^{\star}}}{g_{a^{\star}}'} \right)\left( Q_{M_1^{\star}, M_2^{\star}}' - Q_{M_1^{\star}, M_2^{\star}} \right) \left(\bar{Q}_a' - \bar{Q}_{a^{\star}}' \right) \right\} \\
  &\hspace{0.25in} - P \left\{ \left( \frac{g_{a^{\star}}' - g_{a^{\star}}}{g_{a^{\star}}'}  \right) Q_{M_1^{\star}, M_2^{\star}} \left(\bar{Q}_{a^{\star}}' - \bar{Q}_{a^{\star}} \right) \right\} \\
\end{align*} \endgroup
\noindent The result of the theorem is directly implied by this expression. 

\subsection{Interventional indirect effects} We provide an explicit proof for the interventional effect through $M_1$; the proof for the effect through $M_2$ is nearly identical. We can write $D_{M_1}$ as an estimating function $D_{M_1}(P,\psi_{M_1})$ and \begin{align*}
  P \{D_{M_1}(P', \psi_{M_1})\} &= P \left( \frac{g_a}{g_a'} \left(Q_{M_1, M_2}' - Q_{M_1, M_2} \right)\left[ \left\{\frac{\left(q_{a,M_1}' - q_{a^\star, M_1}'\right)q_{a^\star, M_2}'}{q_{a,M_1,M_2} q_{a,M_1,M_2}'}\right\} \left(\bar{Q}_a' - \bar{Q}_a\right) \right] \right) \\
  &\hspace{0.1in} - P \left\{\left(\frac{g_a' - g_a}{g'_a}\right) Q_{M_1^\star}' Q_{M_2^\star}'\left(\bar{Q}_a' - \bar{Q}_a \right) \right\} \\
  &\hspace{0.1in} - P \left\{\left(\frac{g_{a^\star}' - g_{a^\star}}{g_{a^\star}'}\right) \left( Q_{M_1^\star}'Q_{M_2^\star}' - Q_{M_1^\star}Q_{M_2^\star}\right)\bar{Q}_a \right\} \\
  &\hspace{0.1in} + P \left\{ \left(\frac{g_{a^\star}' - g_{a^\star}}{g_{a^\star}'}\right) Q_{M_1}' \left(Q_{M_2^\star}' - Q_{M_2^\star} \right)\bar{Q}_a'\right\} \\
  &\hspace{0.1in} + P \left\{ \left(\frac{g_a' - g_a}{g'_a}\right) Q_{M_1}'Q_{M_2^\star}' \left(\bar{Q}_a' - \bar{Q}_a \right)\right\} \\
  &\hspace{0.1in} + P \left\{ \left(\frac{g_a' - g_a}{g'_a}\right) (Q_{M_1}' - Q_{M_1})Q_{M_2^\star}' \bar{Q}' \right\} \\
  &\hspace{0.1in} - P \left\{ \left( Q_{M_1}'Q_{M_2^\star}' - Q_{M_1}Q_{M_2^\star} \right) \left( \bar{Q}_a' - \bar{Q}_a \right)\right\} \\
  &\hspace{0.1in} - P \left\{ \left( Q_{M_1}' - Q_{M_1} \right) \left(Q_{M_2^{\star}}' - Q_{M_2^\star} \right) \bar{Q}_a'\right\} \ . 
\end{align*}
\noindent The result of the theorem is directly implied by this expression. 

\section{Proof of asymptotic efficiency}

To prove the joint asymptotic normality of the estimators, it suffices to establish that marginally each estimator is asymptotically linear with influence function equal to the efficient influence function. The joint distribution is then immediately implied. The proof of each marginal estimator is established using results from efficiency theory. Namely, for any pathwise differentiable parameter $\Psi$ of $P$ with gradient $D^*$ and any $P' \in \mathcal{P}$, \begin{align*}
\Psi(P') - \Psi(P) &= (P' - P) D^*(P') + R_2(P, P') \\
&= -P D^*(P') + R_2(P, P') \\
&= (P_n - P) D^*(P') - P_n D^*(P') + R_2(P, P') \\
&= (P_n - P) D^*(P) - P_n D^*(P') + (P_n - P)\{D^*(P') - D^*(P)\} + R_2(P, P') \ .
\end{align*}
Here, $R_2(P, P')$ is the exact second-order remainder and $P_n$ is the empirical measure of $O_1, \dots, O_n$. We can apply this algebra with $P' = P_n'$, where $P_n'$ is any distribution in the model compatible with estimates of the nuisance parameters required to evaluate $D^*$. Note that the one-step estimator is exactly defined as $\Psi(P_n') + P_n D^*(P_n')$, while for targeted minimum loss estimators, by the two-stage construction of the relevant nuisance estimators $P_n D^*(P_n') = o_{\text{p}}(n^{-1/2})$. Thus, proving asymptotic linearity is down to establishing $(P_n - P)\{D^*(P') - D^*(P)\} = o_{\text{p}}(n^{-1/2})$ and $R_2(P, P_n') = o_{\text{p}}(n^{-1/2})$. The former will hold if $D^*(P'_n) - D^*(P)$ falls in a Donsker class with probability tending to one and $P\{D^*(P'_n) - D^*(P)\}^2 = o_{\text{p}}(1)$. The second-order remainder requires more attention. 

First, we recall that in our proof of multiple robustness we established a representation for $P D^*_{\cdot}(P', \psi_{\cdot})$ for each of the effects. Because of the form of each of the efficient influence functions, $D^*_{\cdot}(P', \psi_{\cdot}) = D^*_{\cdot}(P') + \Psi_{\cdot}(P') - \Psi_{\cdot}(P)$ and so $R_{2,\cdot}(P, P') = P D^*_{\cdot}(P', \psi_{\cdot})$. Thus, the terms outlined in our proof of multiple robustness are indeed the exact second-order remainders for their respective parameters. The weakest assumption to prove asymptotic linearity is simply to state that $R_2(P_n', P) = o_{\text{p}}(n^{-1/2})$. However, we may instead prefer to provide conditions on the estimated components of $P_n'$ that are sufficient (but not necessary) to establish $R_2(P_n', P) = o_{\text{p}}(n^{-1/2})$. Indeed, in the simplest cases, we can typically establish rates of convergence of nuisance parameters with respect to $L^2(P)$ norm and then use the Cauchy-Schwarz inequality to argue that the remainder is negligible so long as the product of the rates of the convergence rate of the relevant nuisance parameters is faster than $n^{-1/2}$. In the present problem, proving negligibility of the remainder terms may be more challenging owing to the need to estimate the conditional mediator distribution functions. We can study the third term in the second-order remainder for $\Psi_A$ to illustrate: \begin{align*}
&P \left\{ \frac{g_a}{g_{n,a}} \left(Q_{n,M_1^{\star}, M_2^{\star}} - Q_{M_1^{\star}, M_2^{\star}} \right) \left( \bar{Q}_{n,a} - \bar{Q}_a \right) \right\} \\
&\hspace{0.1in} = \int \left[\frac{g_a(c)}{g'_{n,a}(c)} \int \left(\bar{Q}_{n,a}(m_1, m_2, c) - \bar{Q}_a(m_1, m_2, c)\right) \right. \\ 
&\hspace{1.4in}\left. \times d\left\{Q_{n,a^\star, M_1, M_2}(m_1, m_2 \mid c) - Q_{a^\star, M_1, M_2}(m_1, m_2 \mid c) \right\} \vphantom{\int} \right]dQ_C(c) \ .
\end{align*}
If for each $c$, $Q_{n,a^\star, M_1, M_2}(m_1, m_2 \mid c) - Q_{a^\star, M_1, M_2}(m_1, m_2 \mid c)$ is a right-continuous function with left-hand limits of finite total variation, then the inner integral can be bounded by the product of the sectional variation norm of $Q_{n,a^{\star}, M_1, M_2}(m_1, m_2 \mid c) - Q_{a^{\star}, M_1, M_2}(m_1, m_2 \mid c)$ and the supremum-norm of $Q_{n, a^{\star}, M_1, M_2}(m_1, m_2 \mid c) - Q_{a^{\star}, M_1, M_2}(m_1, m_2 \mid c)$ \citep{van1995efficient}. Thus, we may conclude that the supremum-norm over $c$ of the product of these rates should be faster than $n^{-1/2}$ to achieve the desired negligibility of the second order term. 


\section{Flexible estimators of conditional joint distribution functions}

To estimate the conditional distribution for a given mediator $M_j$, we follow the technique described in \citet{munoz2011super}, which considers estimation of a conditional density via estimation of discrete conditional hazards. Briefly, consider estimation of the distribution of $M_2$ given $A$ and $C$, and, for simplicity, suppose that the support of $M_2$ is $\{1,2,3\}$. We create a long-form data set, where the number of rows contributed by each individual contribute is equal to their observed value of $M_2$. An example is illustrated in Table \ref{long_form_data}. We see that the long-form data set includes an integer-valued column named ``bin'' that indicates to which value of $M_2$ each row corresponds, as well as a binary column $\ind_{\text{bin}}(M_2)$ indicating whether the observed value of $M_2$ corresponds to each bin. These long-form data can be used to fit a regression of the binary outcome $\ind_{\text{bin}}(M_2)$ onto $C$, $A$, and bin. This naturally estimates $\lambda_{b}(a_0, c) = P(M_2 = b \mid M_2 > b-1, A = a_0, C = c)$, the conditional discrete hazard of $M_2$ given $A$ and $C$. Let $\lambda_{n,\cdot}$ denote the estimated hazard obtained from fitting this regression. An estimate of the density at $m_2 \in \mathcal{M}_2$ is \[
q_{n, a_0, M_2}(m_2 \mid c) = \frac{\lambda_{n,m_2}(a_0, c) \prod\limits_{b = 1}^{m_2 - 1} \{1 - \lambda_{n,m_2}(a_0, c)\}}{\sum\limits_{m \in \mathcal{M}_2} \left[\lambda_{n,m}(a_0, c) \prod\limits_{b = 1}^{m - 1} \{1 - \lambda_{n,m}(a_0, c)\} \right]} \ . 
\]
Similarly, an estimate $q_{n,a_0, M_1}(\cdot \mid m_2, c)$ of the conditional distribution of $M_1$ given $A = a_0, M_2 = m_2,C = c$ can be obtained. An estimate of the joint conditional density is thus implied by these estimates, $q_{n, a_0, M_1, M_2}(m_1, m_2 \mid c) = q_{n,a_0, M_1}(m_1 \mid m_2, c) q_{n, a_0, M_2}(m_2 \mid c)$, while an estimate of the marginal distribution of $M_1$ is $q_{n,a_0, M_1}(m_1, \mid c) = \sum_{m_2 \in \mathcal{M}_2} q_{n, a_0, M_1, M_2}(m_1, m_2 \mid c)$.

\begin{table}[ht]
\centering
\begin{tikzpicture}
\node (a) at (0,0){
\begin{tabular}{|cccc|}
\hline
ID  & $C$ & $A$ & $M_2$ \\ \hline
1 & 1   & 0 & 1   \\
2 & 0   & 1 & 3   \\
\vdots & \vdots & \vdots & \vdots \\
$n$ & 0   & 1 & 2   \\ 
\hline
\end{tabular}};
\node[xshift=6cm] (b)
{
\begin{tabular}{|ccccc|}
\hline
ID  & $C$ & $A$ & bin & $\ind_{\text{bin}}(M_2)$ \\ \hline
1 & 1   & 0 & 1   & 1 \\
2 & 0   & 1 & 1   & 0 \\
2 & 0   & 1 & 2   & 0 \\
2 & 0   & 1 & 3   & 1 \\
\vdots & \vdots & \vdots & \vdots & \vdots \\
$n$ & 0   & 1 & 1   & 0 \\
$n$ & 0   & 1 & 2   & 1 \\
\hline
\end{tabular}
};
\draw[->,ultra thick](a)--(b);
\end{tikzpicture}
\caption{An illustration of how to make a long form data set suitable for estimating mediator distributions. An ID is uniquely assigned to each independent data unit and a single confounder $C$ is included in the mock data set.}
\label{long_form_data}
\end{table}

\section{Details of targeted minimum loss-based estimation}

Targeted minimum loss-based estimation is a two step estimation process. In the first step, estimates of key nuisance parameters are generated. The second step involves finding a \emph{parametric submodel} and a \emph{loss function} that can be used in a process referred to as \emph{targeting} the nuisance parameter estimates. The goal of this targeting step is to map the initial estimates of nuisance parameters into a revised estimate that simultaneously (i) are no worse at estimating their true counterparts, and (ii) solve a set of user-specified equations. This is achieved by, possibly iterative, empirical risk minimization along a low-dimensional parametric model for the nuisance parameters. Such a submodel is index by a parameter $\epsilon$ and is arranged so that setting $\epsilon = 0$ returns the initial estimate. Risk minimization along such a submodel generally ensures goal (i), while goal (ii) is assured by ensuring that the user-specified equations, at least approximately, span the score of $\epsilon$ at $\epsilon = 0$. Early examples of this approach appeared in \citet{Scharfstein:Rotnitzky:Robins99} and \citet{Bang:Robins05}, while the first general treatment was presented in \citet{vanderLaan:Rubin06}. A short overview appears in \citet{van2018targeted} and comprehensive treatments are provided in \citet{vanderLaan:Rose11} and \citet{van2018targetedbook}. 

From a high level, our targeted minimum loss estimator of the vector $(\psi, \psi_A, \psi_{M_1}, \psi_{M_2}, \psi_{M_1,M_2})$ is implemented in the following way. Given initial estimates $\bar{Q}_{n,a_0}$ of the outcome regression and $g_{n,a_0}$ of the propensity score, we define a logistic regression submodel for $\bar{Q}_{n,a_0}$. This submodel contains parameters that generate score equations that appear across the efficient influence functions of $\psi_A, \psi_{M_1},$ and $\psi_{M_2})$. We denote by $\bar{Q}_{n,\cdot}^*$ the targeted outcome regression estimate. Initial estimates of the mediator-marginalized $\tilde{Q}$ parameters are generated by marginalizing $\bar{Q}_{n,\cdot}^*$ with respect to the initial estimators of the relevant mediator distributions. These mediator-marginalized parameters are then themselves targeted using a separate submodel for each of component of $(\psi, \psi_A, \psi_{M_1}, \psi_{M_2})$. Finally, the effect estimate is given by marginalizing with respect to the empirical distribution of $C$. 

Specifically, our estimator may be implemented in the following steps. Define the following univariate logistic regression model for the conditional mean outcome given $A = a_0, M_1 = m_1, M_2 = m_2,$ and $C = c$, \[
 \bar{Q}_{n,a_0, \epsilon}(m_1, m_2, c) = \mbox{expit}[\mbox{logit}\{\bar{Q}_{n,a_0}(m_1, m_2, c)\} + \epsilon^{\T} H_{n,a_0}(m_1, m_2, c)] \ , \ \epsilon \in \mathbb{R} \ ,
\]
where $\epsilon = (\epsilon_1, \dots, \epsilon_5)^{\T}$ and $H_{n,\cdot} = (H_{1,n,\cdot}, \dots, H_{5,n,\cdot})^{\T}$ where \begin{align*}
H_{1,n,\tilde{a}}(\tilde{m}_1, \tilde{m}_2, \tilde{c}) &= \frac{\ind_{a}(\tilde{a})}{g_{n,a}(\tilde{c})} \frac{q_{n,a^\star,M_1,M_2}(\tilde{m}_1, \tilde{m}_2 \mid \tilde{c})}{q_{n, a, M_1,M_2}(\tilde{m}_1, \tilde{m}_2 \mid \tilde{c})} \\ 
H_{2,n,\tilde{a}}(\tilde{c}) &= \frac{\ind_{a^\star}(\tilde{a})}{g_{n,a^\star}(\tilde{c})} \\
H_{3,n,\tilde{a}}(\tilde{m}_1, \tilde{m}_2, \tilde{c}) &= \frac{\ind_{a}(\tilde{a})}{g_{n,a}(\tilde{c})} \frac{q_{n, a, M_1}(\tilde{m}_1 \mid \tilde{c})q_{n,a^\star, M_2}(\tilde{m}_2 \mid \tilde{c})}{q_{n, a, M_1,M_2}(\tilde{m}_1, \tilde{m}_2 \mid \tilde{c})} \\ 
H_{4,n,\tilde{a}}(\tilde{m}_1, \tilde{m}_2, \tilde{c}) &= \frac{\ind_{a}(\tilde{a})}{g_{n,a}(\tilde{c})} \frac{q_{n,a^\star, M_1}(\tilde{m}_1 \mid \tilde{c})q_{n,a^\star, M_2}(\tilde{m}_2 \mid \tilde{c})}{q_{n, a, M_1,M_2}(\tilde{m}_1, \tilde{m}_2 \mid \tilde{c})} \\ 
H_{5,n,\tilde{a}}(\tilde{m}_1, \tilde{m}_2, \tilde{c}) &= \frac{\ind_{a}(\tilde{a})}{g_{n,a}(\tilde{c})} \frac{q_{n,a, M_1}(\tilde{m}_1 \mid \tilde{c})q_{n,a, M_2}(\tilde{m}_2 \mid \tilde{c})}{q_{n, a, M_1,M_2}(\tilde{m}_1, \tilde{m}_2 \mid \tilde{c})} 
\end{align*}

The parameter $\epsilon$ of this regression model is estimated via maximum (quasi-)likelihood. This can be achieved using optimization routines included in standard regression software packages, such as via iteratively re-weighted least squares as in the \texttt{glm} function in the \texttt{R} programming language. Denoting by $\epsilon_n$ the estimated value of $\epsilon$, we define $\bar{Q}_{n,a_0}^* = \bar{Q}_{n,a_0,\epsilon_n}$ to be our targeted outcome regression estimate. 

We then proceed for the various effects as follows:
\subsection{Direct effect} We marginalize $\bar{Q}_{n,a_0}^*$ using the initial estimate of the joint conditional mediator distribution, as in equation (2) of the main paper; however, rather than using the initial outcome regression estimate, we use the revised estimate $\bar{Q}_{n,a_0}^*$. With an abuse of notation, we use $\tilde{Q}_{n,a_0, M_1^{\star}, M_2^{\star}}$ to denote these estimates. Next, we define a logistic regression model for the \emph{difference} in conditional interventional direct effects scaled to the unit-interval. Specifically, we define $\bar{Q}_{\Delta,\text{scaled}} = (\bar{Q}_{a} - \bar{Q}_{a^{\star}} + 1)/2$ and note that $\bar{Q}_{\Delta,\text{scaled}}$ maps $(m_1, m_2, c)$ to the unit interval. A natural estimate of this quantity is $\bar{Q}_{n,\Delta, \text{scaled}} = (\bar{Q}_{n,a}^* - \bar{Q}_{n,a}^* + 1)/2$. Similarly, we define $\tilde{Q}_{\Delta,\text{scaled}} = (\tilde{Q}_{a, M_1^{\star}, M_2^{\star}} - \tilde{Q}_{a^{\star}, M_1^{\star}, M_2^{\star}} + 1)/2$ so that $\tilde{Q}_{\Delta,\text{scaled}}$ is simply the conditional mean of $\bar{Q}_{\Delta, \text{scaled}}$ with respect to the joint mediator distribution given $A = a^\star, C$. Similarly as above, a natural estimate of this quantity is $\tilde{Q}_{n,\Delta, \text{scaled}} = (\tilde{Q}_{n, a, M_1^{\star}, M_2^{\star}} - \tilde{Q}_{n, a^{\star}, M_1^{\star}, M_2^{\star}} + 1)/2$. Given these estimates, we define the following logistic regression model for $\tilde{Q}_{\Delta,\text{scaled}}$, \[
  \tilde{Q}_{\Delta,\text{scaled}, \delta} = \mbox{expit}[\mbox{logit}\{\tilde{Q}_{n,\Delta, \text{scaled}}\} + \delta/g_{n,a^{\star}}] \ , \ \delta \in \mathbb{R} \ . 
\]
The single parameter $\delta$ of this regression model can be estimated by regressing the estimated outcome $\bar{Q}_{n,\Delta, \text{scaled}}$ onto the single covariate $1/g_{n,a^{\star}}$ amongst observations with $A = a^{\star}$. Let $\delta_n$ denote the estimated value of $\delta$, and define $\tilde{Q}_{n,\Delta}^* = 2\tilde{Q}_{\Delta,\text{scaled}, \delta_n} - 1$ to be the revised estimate of the conditional interventional direct effect. Note that $\tilde{Q}_{n,\Delta}^*$ maps $c$ to $(-1, 1)$. Finally, we marginalize this distribution with respect to the empirical distribution of $C$, \[
  \psi_{n,A}^* = \int \tilde{Q}_{n,\Delta}^*(c) dQ_{n,C}(c) = n^{-1}\sum\limits_{i=1}^n \tilde{Q}_{n,\Delta}^*(C_i) \ . 
\]

\subsection{Indirect effect} We describe the procedure for the indirect effect through $M_1$; the effect through $M_2$ is similar. We marginalize $\bar{Q}_{n,a_0}^*$ using the initial estimates of the marginal mediator distributions and define \begin{align*}
\tilde{Q}_{n, a, M_1 \times M_2^{\star}}(c) &= \int\limits_{\mathcal{M}_1} \int\limits_{\mathcal{M}_2} \bar{Q}_{n,a}^*(m_1, m_2, c) dQ_{n,a, M_1}(m_1 \mid c) dQ_{n,a^\star,M_2}(m_2 \mid c) \ , \ \mbox{and} \\
\tilde{Q}_{n, a, M_1^{\star} \times M_2^{\star}}(c) &= \int\limits_{\mathcal{M}_1} \int\limits_{\mathcal{M}_2} \bar{Q}_{n,a}^*(m_1, m_2, c) dQ_{n,a^\star, M_1}(m_1 \mid c) dQ_{n,a^\star,M_2}(m_2 \mid c) \ .
\end{align*}
Selecting appropriate submodels for these quantities turns out to be an interesting problem. The problem has two interesting features. First, due to Fubini's theorem, the parameter $\tilde{Q}_{a, M_1 \times M_2^\star}$ can be viewed as (a) the conditional mean of $\tilde{Q}_{a, M_1}(M_2, C)$ given $C$ with respect to the marginal distribution $M_2$ given $A = a^\star, C$, as well as (b) the conditional mean of $\tilde{Q}_{a, M_2^\star}(M_1,C)$ given $C$ with respect to the marginal distribution of $M_1$ given $A = a, C$. Similarly, for $\tilde{Q}_{n, a, M_1^{\star} \times M_2^{\star}}$. The natural inclination then is to define a sum loss function. Specifically, we define \begin{align*}
&\mathcal{L}(\tilde{Q}_{a,M_1 \times M_2^\star}' \mid \tilde{Q}_{a, M_1}', \tilde{Q}_{a, M_2^\star}', g_a', g_{a^\star}')(\tilde{o}) \\
&\hspace{0.05in} = -\left(\frac{\ind_{a}(\tilde{a})}{g_a'(\tilde{c})} \left[ \tilde{Q}_{a, M_1}'(\tilde{m}_2, \tilde{c}) \mbox{log}\{\tilde{Q}_{a, M_1 \times M_2^\star}'(\tilde{c})\} + \{1 - \tilde{Q}_{a, M_1}'(\tilde{m}_2, \tilde{c})\} \mbox{log}\{1 - \tilde{Q}_{a, M_1 \times M_2^\star}'(\tilde{c})\}   \right] \right. \\
&\hspace{0.21in} \left. + \frac{\ind_{a^\star}(\tilde{a})}{g_{a^\star}'(\tilde{c})} \left[ \tilde{Q}_{a, M_2^\star}'(\tilde{m}_1, \tilde{c}) \mbox{log}\{\tilde{Q}_{a, M_1 \times M_2^\star}'(\tilde{c})\} + \{1 - \tilde{Q}_{a, M_2^\star}'(\tilde{m}_1, \tilde{c})\} \mbox{log}\{1 - \tilde{Q}_{a, M_1 \times M_2^\star}'(\tilde{c})\}   \right] \right) \ . 
\end{align*}
We note that this loss is indexed by the nuisance parameters $\tilde{Q}_{a, M_1}', \tilde{Q}_{a, M_2^\star}', g_a', g_{a^\star}'$. We can use an intercept-only logistic regression submodel $\tilde{Q}_{n, a, M_1 \times M_2^{\star}, \eta} = \mbox{expit}[\mbox{logit}\{\tilde{Q}_{n,a, M_1 \times M_2^{\star}}\} + \eta] \ , \ \eta \in \mathbb{R}$. Let $\eta_n = \mbox{argmin}_{\eta \in \mathbb{R}} P_n \mathcal{L}(\tilde{Q}_{n,a,M_1 \times M_2^\star,\eta} \mid \tilde{Q}_{n,a, M_1}, \tilde{Q}_{n,a, M_2^\star}, g_{n,a}, g_{n,a^\star})$ be the minimum loss estimator of $\eta$ and define $\tilde{Q}_{n, a, M_1 \times M_2^{\star}}^* = \tilde{Q}_{n, a, M_1 \times M_2^{\star}, \eta_n}$. Standard software can be used to perform this risk minimization. Specifically, we can fit an intercept-only logistic regression where the outcome of the regression is $\tilde{Q}_{n, a, M_1}'(M_{2i}, C_i)$ for observations with $A_i = a$ and $\tilde{Q}_{n, a, M_2^\star}'(M_{1i}, C_i)$ for observations with $A_i = a^\star$. We include a vector of weights in this procedure as well with weights equal to $1/g_{n,a}(C_i)$ if $A_i = a$ and $1/g_{n,a^\star}(C_i)$ if $A_i = a^\star$.  

Similarly, to target $\tilde{Q}_{n, a, M_1^{\star} \times M_2^{\star}}$, we define the sum loss function \begin{align*}
&\mathcal{L}(\tilde{Q}_{a,M_1^\star \times M_2^\star}' \mid \tilde{Q}_{a, M_1^\star}', \tilde{Q}_{a, M_2^\star}', g_a', g_{a^\star}')(\tilde{o}) \\
&\hspace{0.05in} = -\left(\frac{\ind_{a}(\tilde{a})}{g_a'(\tilde{c})} \left[ \tilde{Q}_{a, M_1^\star}'(\tilde{m}_2, \tilde{c}) \mbox{log}\{\tilde{Q}_{a, M_1^\star \times M_2^\star}'(\tilde{c})\} + \{1 - \tilde{Q}_{a, M_1^\star}'(\tilde{m}_2, \tilde{c})\} \mbox{log}\{1 - \tilde{Q}_{a, M_1^\star \times M_2^\star}'(\tilde{c})\}   \right] \right. \\
&\hspace{0.21in} \left. + \frac{\ind_{a^\star}(\tilde{a})}{g_{a^\star}'(\tilde{c})} \left[ \tilde{Q}_{a, M_2^\star}'(\tilde{m}_1, \tilde{c}) \mbox{log}\{\tilde{Q}_{a, M_1^\star \times M_2^\star}'(\tilde{c})\} + \{1 - \tilde{Q}_{a, M_2^\star}'(\tilde{m}_1, \tilde{c})\} \mbox{log}\{1 - \tilde{Q}_{a, M_1^\star \times M_2^\star}'(\tilde{c})\}   \right] \right) \ , 
\end{align*}
and the submodel $\tilde{Q}_{n, a, M_1^\star \times M_2^{\star}, \gamma} = \mbox{expit}[\mbox{logit}\{\tilde{Q}_{n,a, M_1^\star \times M_2^{\star}}\} + \gamma] \ , \ \gamma \in \mathbb{R}$. Let $\gamma_n$ be the minimum loss estimator of $\gamma$. As above, standard software can be used to perform this risk minimization with obvious modifications to the procedure outlined about for targeting $\tilde{Q}_{n, a, M_1 \times M_2^{\star}}$. 
Denote the targeted estimate as $\tilde{Q}_{n, a, M_1^\star \times M_2^{\star}}^* = \tilde{Q}_{n, a, M_1^\star \times M_2^{\star}, \gamma_n}$.

The final estimate is \[
\psi_{n,M_1}^* = \int \left\{\tilde{Q}_{n, a, M_1 \times M_2^{\star}}^*(c) - \tilde{Q}_{n, a, M_1^\star \times M_2^{\star}}^*(c) \right\} dQ_{n,C} \ . 
\]

\subsection{Total effect} Since we generate our estimate of the covariant effect as the difference in an estimate of the total effect and the sum of the direct and indirect effects, we require a targeting procedure for the total effect. For this, we marginalize $\bar{Q}_{n,a_0}^*$ with respect to the initial estimate of the joint mediator distribution $Q_{n,a_0,M_1, M_2}(\cdot, \cdot \mid c)$, giving us initial estimates $\tilde{Q}_{n,a,M_1,M_2}$ and $\tilde{Q}_{n,a^\star,M_1^\star,M_2^\star}$ of $\tilde{Q}_{a,M_1,M_2}$ and $\tilde{Q}_{a^\star,M_1^\star,M_2^\star}$, respectively. These quantities can then be used in a standard targeted minimum loss estimator procedure, as described in e.g., \citet{vanderLaan:Rose11}.

\section{Extensions}

\emph{Other effect scales:} We illustrate how the indirect effect through $M_1$ can be cast as a multiplicative effect. To that end, note that the efficient influence function of the parameter $\Psi_{M_1, a}(P') = \int \tilde{Q}_{a, M_1 \times M_2^\star}(c) dQ_C(c)$ is \begin{align*}
  D_{M_1, a}^*(P')(\tilde{o}) &= \frac{\ind_a(\tilde{a})}{g'_a(\tilde{c})} \frac{q_{a, M_1}'(\tilde{m}_1 \mid \tilde{c})q_{a^\star, M_2}'(\tilde{m}_2 \mid \tilde{c})}{q_{a, M_1,M_2}'(\tilde{m}_1, \tilde{m}_2 \mid \tilde{c})} \{\tilde{y} - \bar{Q}_a'(\tilde{m}_1, \tilde{m}_2, \tilde{c})\} \\
&\hspace{0.2in} + \frac{\ind_a(\tilde{a})}{g'_a(\tilde{c})} \{\tilde{Q}_{a, M_2^{\star}}'(\tilde{m}_1, \tilde{c}) - \tilde{Q}_{a, M_1 \times M_2^\star}'(\tilde{c})\} \\
&\hspace{0.3in} + \frac{\ind_{a^{\star}}(\tilde{a})}{g'_{a^\star}(\tilde{c})} \left\{\tilde{Q}_{a, M_1}'(\tilde{m}_2, \tilde{c}) - \tilde{Q}_{a, M_1 \times M_2^\star}'(\tilde{c}) \}\right\} \\
&\hspace{0.5in} + \tilde{Q}_{a, M_1 \times M_2^\star}'(\tilde{c}) - \Psi_{M_1,a}(P') \ .
\end{align*}
Similarly, the efficient influence function of $\Psi_{M_1, a^\star}(P') = \int \tilde{Q}_{a, M_1^\star \times M_2^\star}(c) dQ_C(c)$ is \begin{align*}
D_{M_1, a^\star}^*(P')(\tilde{o}) &= \frac{\ind_a(\tilde{a})}{g'_a(\tilde{c})} \frac{q_{a^\star, M_1}'(\tilde{m}_1 \mid \tilde{c})q_{a^\star, M_2}'(\tilde{m}_2 \mid \tilde{c})}{q_{a, M_1,M_2}'(\tilde{m}_1, \tilde{m}_2 \mid \tilde{c})} \{\tilde{y} - \bar{Q}_a'(\tilde{m}_1, \tilde{m}_2, \tilde{c})\} \\
&\hspace{0.3in} + \frac{\ind_{a^{\star}}(\tilde{a})}{g'_{a^{\star}}(\tilde{c})} \{\tilde{Q}_{a, M_2^{\star}}'(\tilde{m}_1, \tilde{c}) - \tilde{Q}_{a, M_1^\star \times M_2^\star}'(\tilde{c})\} \\
&\hspace{0.4in} + \frac{\ind_{a^{\star}}(\tilde{a})}{g'_{a^\star}(\tilde{c})} \left\{\tilde{Q}_{a, M_1^{\star}}'(\tilde{m}_2, \tilde{c}) - \tilde{Q}_{a, M_1^{\star} \times M_2^{\star}}'(\tilde{c})\right\} \\
&\hspace{0.5in} + \tilde{Q}_{a, M_1^\star \times M_2^\star}'(\tilde{c}) - \Psi_{M_1,a^\star}(P') \ .
\end{align*}
It is no surprise that $D^*_{M_1} = D^*_{M_1, a} - D^*_{M_1, a^\star}$. A one-step estimator of $\psi_{M_1,a_0} = \Psi_{M_1,a_0}(P)$ is $\psi_{n,M_1,a_0} = \Psi_{M_1, a_0}(P_n') + n^{-1} \sum_{i=1}^n D^*_{M_1,a_0}(P_n')(O_i)$, while a targeted minimum loss estimator can be generated by straightforward modifications to the procedure described above. 

Suppose we desire to make inference on the ratio scale, $\psi_{M_1, a} / \psi_{M_1, a^\star}$. Define the vector influence function $D^*_{M_1,\cdot} = (D^*_{M_1, a}, D^*_{M_1,a^{\star}})$. By the central limit theorem the scaled vector $n^{1/2}(\psi_{n,M_1,a}, \psi_{n,M_1,a^\star})$ converges in distribution to a bivariate Normal random variable with mean $(\psi_{M_1,a}, \psi_{M_1,a^\star})$ and covariance matrix $\Sigma_{M_1} = \int D^*_{M_1, \cdot}(P)(o) D^*_{M_1, \cdot}(o)^\T dP(o)$. The delta method then implies that the scaled estimator $n^{1/2}(\psi_{n,M_1,a}/\psi_{n,M_1,a^\star})$ converges in distribution to a Normal random variable with mean $(\psi_{M_1,a}/\psi_{M_1,a^\star})$ and covariance $$\tau^2 = (1 / \psi_{M_1,a^\star}, -\psi_{M_1,a^\star}/\psi_{M_1,a^\star}^2) \Sigma_{M_1} (1 / \psi_{M_1,a^\star}, -\psi_{M_1,a^\star}/\psi_{M_1,a^\star}^2)^\T \ . $$
A natural estimator of $\tau$ would plug in one-step estimators of $\psi_{M_1, a_0}$ and the empirical covariance matrix of $D^*_{M_1,\cdot}(P_n')$ to this formula.

\emph{More than two mediators:} We provide the form of the efficient influence function for the interventional direct and indirect effect with multiple mediators. For the direct effect, the extension is straightforward: examination of the proof for the efficient influence function with only two mediators reveals that the result immediately generalizes to higher dimensional mediators. We introduce the shorthand $M_{1:t} = (M_1, M_2, \dots, M_t)$, and define $Q_{M_{1:t}}(m_{1:t} \mid a_0, c)$ as the joint cumulative distribution of $M_{1:t}$ given $A = a_0, C = c$ evaluated at $m_{1:t}$. Similarly, we denote by $\bar{Q}_{a_0}(m_{1:t}, c)$ the conditional mean of $Y$ given $M_{1:t} = m_{1:t}, C = c$. We define \begin{align*}
  \tilde{Q}_{a^{\star}, M_{1:t}^\star}(c) &= \int\limits_{\mathcal{M}_1 \times \ldots \times \mathcal{M}_t} \bar{Q}_{a^{\star}}(m_{1:t}, c) \ dQ_{a^\star, M_{1:t}}(m_{1:t} \mid c) \ \mbox{and} \\
  \tilde{Q}_{a, M_{1:t}^{\star}}(c) &= \int\limits_{\mathcal{M}_1 \times \ldots \times \mathcal{M}_t} \bar{Q}_{a}(m_{1:t}, c) \ dQ_{a^\star, M_{1:t}}(m_{1:t} \mid c) \ . 
\end{align*}
The efficient influence function of the direct effect under sampling from $P' \in \mathcal{P}$ is \begin{align*}
D^*_{t,A}(P')(\tilde{o}) &= \frac{\ind_a(\tilde{a})}{g'_a(\tilde{c})} \frac{q_{a^\star, M_{1:t}}'(\tilde{m}_{1:t} \mid \tilde{c})}{q_{a, M_{1:t}}'(\tilde{m}_{1:t} \mid \tilde{c})} \{\tilde{y} - \bar{Q}_a'(\tilde{m}_{1:t}, \tilde{c})\} \\
&\hspace{0.2in} - \frac{\ind_{a^{\star}}(\tilde{a})}{g'_{a^\star}(\tilde{c})} \{\tilde{y} - \bar{Q}_{a^\star}'(\tilde{m}_{1:t}, \tilde{c})\} \\
&\hspace{0.3in} + \frac{\ind_{a^{\star}}(\tilde{a})}{g'_{a^\star}(\tilde{c})} \left[\bar{Q}_a'(\tilde{m}_{1:t}, \tilde{c}) - \bar{Q}_{a^\star}'(\tilde{m}_{1:t}, \tilde{c}) - \{\tilde{Q}'_{a, M_{1:t}^\star}(\tilde{c}) - \tilde{Q}_{a^{\star}, M_{1:t}^\star}'(\tilde{c})\}\right] \\ 
&\hspace{0.4in} + \tilde{Q}_{a, M_{1:t}^\star}'(\tilde{c}) - \tilde{Q}_{a^{\star},M_{1:t}^\star}'(\tilde{c}) - \Psi_{t,A}(P') \ . 
\end{align*}

The efficient influence functions for indirect effects require more effort to derive. The proof is largely similar to the case of two mediators though, and so is omitted here. We require additional notation. For $s = 1,\dots,t$, let $q_{M_{\bar{s}}}(m_{\bar{s}} \mid c) = \prod_{u=1}^{s-1} q_{a,M_u}(m_u \mid c) \prod_{v=s+1}^{t} q_{a^\star, M_v}(m_v \mid c)$ denote the product of all marginal mediator densities besides that for $M_s$, where the mediator densities are conditional on $A = a, C$ for mediators $M_1, \dots, M_{s-1}$ and on $A = a^\star, C$ for mediators $M_{s+1}, \dots, M_t$. We use a natural extension of our notation above to denote the outcome regression marginalized with respect to the product of marginal mediator distributions. For example, \begin{align*}
  \tilde{Q}_{a, \prod_{u=1}^{s-1} M_{u} \times \prod_{v=s+1}^t M_{v}^\star}(m_s, c) &= \int \bar{Q}_a(m_{1:t}, c) dQ_{M_{\bar{s}}}(m_{\bar{s}} \mid c) \ .
\end{align*}
The efficient influence function of the indirect effect through $M_s, s = 1,\dots,t$ under sampling from $P' \in \mathcal{P}$ is \begin{align*}
D^*_{t, M_s}(P')(\tilde{o}) &= \frac{\ind_a(\tilde{a})}{g'_a(\tilde{c})} \frac{\{q_{a, M_s}'(\tilde{m}_s \mid \tilde{c}) - q_{a^\star, M_s}'(\tilde{m}_s \mid \tilde{c})\}q_{M_{\bar{s}}}'(m_{\bar{s}} \mid \tilde{c})}{q_{a, M_{1:t}}'(\tilde{m}_{1:t} \mid \tilde{c})} \{\tilde{y} - \bar{Q}_a'(\tilde{m}_{1:t}, \tilde{c})\} \\
&\hspace{0.2in} + \frac{\ind_a(\tilde{a})}{g'_a(\tilde{c})} \{\tilde{Q}_{a, \prod_{u=1}^{s-1} M_{u} \times \prod_{v=s+1}^t M_{v}^\star}(\tilde{m}_s, \tilde{c}) - \tilde{Q}_{a, \prod_{u=1}^s M_{u} \times \prod_{v=s+1}^t M_{v}^\star}(\tilde{c}) \} \\
&\hspace{0.2in} - \frac{\ind_a(\tilde{a})}{g'_a(\tilde{c})} [\tilde{Q}_{a, M_s \times \prod_{v=s+1}^{t} M_{v}^\star}(\tilde{m}_{1:s-1}, \tilde{c}) - \tilde{Q}_{a, \prod_{v=s}^{t} M_{v}^\star}(\tilde{m}_{1:s-1}, \tilde{c}) \\
&\hspace{1.0in} - \{\tilde{Q}_{a, \prod_{u=1}^{s} M_{u} \times \prod_{v=s+1}^t M_{v}^\star}(\tilde{c}) - \tilde{Q}_{a, \prod_{u=1}^{s-1} M_{u} \times \prod_{v=s}^t M_{v}^\star}(\tilde{c}) \}] \\
&\hspace{0.2in} - \frac{\ind_{a^\star}(\tilde{a})}{g'_{a^{\star}}(\tilde{c})} \{\tilde{Q}_{a, \prod_{u=1}^{s-1} M_{u} \times \prod_{v=s+1}^t M_{v}^\star}(\tilde{m}_s, \tilde{c}) - \tilde{Q}_{a, \prod_{u=1}^{s-1} M_{u} \times \prod_{v=s}^t M_{v}^\star}(\tilde{c}) \} \\
&\hspace{0.2in} + \frac{\ind_{a^\star}(\tilde{a})}{g'_{a^{\star}}(\tilde{c})} [\tilde{Q}_{a, \prod_{u=1}^{s} M_{u}}(\tilde{m}_{s+1:t}, \tilde{c}) - \tilde{Q}_{a, \prod_{u=1}^{s-1} M_{u} \times M_s^\star} (\tilde{m}_{s+1:t}, \tilde{c}) \\
&\hspace{1.0in} - \{\tilde{Q}_{a, \prod_{u=1}^{s} M_{u} \times \prod_{v=s+1}^t M_{v}^\star}(\tilde{c}) - \tilde{Q}_{a, \prod_{u=1}^{s-1} M_{u} \times \prod_{v=s}^t M_{v}^\star}(\tilde{c}) \}] \\
&\hspace{0.2in} + \tilde{Q}_{a, \prod_{u=1}^{s} M_{u} \times \prod_{v=s+1}^t M_{v}^\star}(\tilde{c}) - \tilde{Q}_{a, \prod_{u=1}^{s-1} M_{u} \times \prod_{v=s}^t M_{v}^\star}(\tilde{c}) - \Psi_{t, M_s}(P') \ . 
\end{align*}

\end{document}